\documentclass[aps,prb,reprint,twocolumn,groupedaddress,superscriptaddress,amsmath,amssymb,amsfonts,floats,floatfix,noraggedbottom,balancelastpage,dvips,showpacs,10pt]{revtex4-1}

\usepackage{epsfig} 
\usepackage{psfrag}
\usepackage{braket}
\usepackage{color}
\usepackage[usenames,dvipsnames]{xcolor}
\usepackage{textcomp}
\usepackage{graphicx}
\usepackage{subfigure}
\usepackage{latexsym}
\usepackage{hyperref}

\hypersetup{dvips,
  pdfauthor={Qingyou Meng, Christopher N. Varney, Hans Fangohr, Egor Babaev},
  pdftitle={Phase diagrams of vortex matter with multi-scale
    inter-vortex interactions in layered superconductors},
  letterpaper,
  colorlinks=true,
  linkcolor=blue,
  citecolor=blue,
}

\usepackage{breakurl}

\begin{document}

\title{Phase diagrams of vortex matter with multi-scale inter-vortex
  interactions in layered superconductors}

\author{Qingyou Meng}
\affiliation{Department of Physics, University of Massachusetts, Amherst,
  Massachusetts 01003, USA}
\author{Christopher N. Varney}
\affiliation{Department of Physics, University of West Florida,
  Pensacola, FL 32514, USA}
\author{Hans Fangohr}
\affiliation{Engineering and the Environment, University of
  Southampton, SO17 1BJ, UK}
\author{Egor Babaev}  
\affiliation{Department of Theoretical Physics, The Royal Institute of
  Technology, SE-10691 Stockholm, Sweden}

\begin{abstract}
  It was recently proposed to use the stray magnetic fields of
  superconducting vortex lattices to trap ultracold atoms for building
  quantum emulators. This calls for new methods for engineering and
  manipulating of the vortex states. One of the possible routes
  utilizes type-1.5 superconducting layered systems with multi-scale
  inter-vortex interactions. In order to explore the possible vortex
  states that can be engineered, we present two phase diagrams of
  phenomenological vortex matter models with multi-scale inter-vortex
  interactions featuring several attractive and repulsive length
  scales. The phase diagrams exhibit a plethora of phases, including conventional 2D lattice
  phases, five stripe phases, dimer, trimer, and tetramer phases, void
  phases, and stable low-temperature disordered phases. The transitions between these states can
  be controlled by the value of an applied external field.
\end{abstract}


\maketitle

%%%%%%%%%%%%%%%%%%%%%%%  Main text of paper  %%%%%%%%%%%%%%%%%%%%%%%%%
\section{\label{sec:introduction}Introduction}

Recently much interest was generated by the proposal of quantum
systems built by placing ultracold atoms in a lattice potential
generated via stray magnetic fields of superconducting
vortices.~\cite{romero-isart_superconducting_2013} This gives rise to
a possibility of creating quantum emulators. To this end, better
control of vortex lattices in superconductors is
required. Predominantly vortex lattices in superconductors have
hexagonal symmetry, the square lattices are possible but
rare.~\cite{aegerter_evidence_1998,riseman_observation_1998,ray_muon-spin_2014}

One route for creating more complicated vortex lattices is through
pinning the vortices by artificial pinning
centers.~\cite{moshchalkov_nanoscience_2010,baert_composite_1995,moshchalkov_magnetization_1996,rosseel_depinning_1996, grigorenko_direct_2001,grigorenko_symmetry_2003,berdiyorov_vortex_2006}
This approach, demonstrated by Romero {\em et al.} in
Ref.~\onlinecite{romero-isart_superconducting_2013}, appears
technically challenging to achieve the requirements for perfections of
the vortex lattice due to possible variations and field
inhomogeneities in the anti-dot arrays.
% In Romero {\em et al.}~\cite{romero-isart_superconducting_2013}, the
% technical difficulties with this approach requirements for
% perfection of the vortex lattice and possible variations and field
% inhomogeneities in the anti-dot arrays.

In Ref.~\onlinecite{meng_honeycomb_2014}, an alternative approach was
proposed that involves multi-component superconducting type-1.5
superconducting systems. In such systems, there are several coherence
lengths that can be smaller and/or larger than the penetration length  (we absorb 
a factor of ${2}^{-1/2}$ in the definition of coherence length):
$\xi_1, \xi_2, \ldots < \lambda < \xi_i, \ldots,
\xi_N$.~\cite{babaev_semi-meissner_2005, silaev_microscopic_2011, silaev_microscopic_2012, garaud_vortex_2012, PhysRevB.84.134518,  PhysRevB.84.134515, PhysRevLett.105.067003} 
 The first claim of experimental observation of this behavior was
originally reported in  works on MgB$_2$
\cite{moshchalkov_type-1.5_2009, gutierrez_scanning_2012,dao_giant_2011}. Recently a similar 
claim   was reported  in experimental studies of
Sr$_2$RuO$_4$ ~\cite{hicks_limits_2010, ray_muon-spin_2014}   and $\rm LaPt_3 Si$ \cite{noncentr,1347-4065-54-4-048001}. 
The
non-monotonic inter-vortex interaction is also possible in
electromagnetically or proximity-effect-coupled
bilayers.~\cite{babaev_semi-meissner_2005}
  
In case of two-component superconductors, the long-range inter-vortex
interaction energy is given by~\cite{babaev_semi-meissner_2005,  carlstrom_type-1.5_2011,silaev_microscopic_2011}
\begin{align}
  E_{\rm int}=C_B^2K_0\left( \frac{r}{\lambda}\right) - C_1^2
  K_0 \left(\frac{r}{\xi_1}\right) - C_2^2K_0
  \left(\frac{r}{\xi_2}\right)
  \label{eq:inn}
\end{align}
Here, the first term describes inter-vortex repulsion which comes from
magnetic and current-current interaction and has the length scale of
magnetic field penetration length. The second and third terms
describes attractive interactions from cores overlaps with ranges set
by coherence lengths. %In Ref.~\onlinecite{varney_hierarchical_2013}, a
%further generalization was discussed where the interaction has
 Multiple repulsive length scales can arise in layered systems due
  to different penetration lengths in different layers or due to
  splaying of magnetic field in interlayer spacing of
  superconductor-insulator-superconductor systems.
% For a straight and rigid vortex
%line, the asymptotical long-range interaction is then
%\begin{align}
%  E_{\rm int} = \sum_i C_B{}_i^2K_0\left( \frac{r}{\lambda_i}\right) -
 % \sum_jC_j^2{2\pi}K_0\left(\frac{r}{\xi_j}\right).
%\label{eq:inn}
%\end{align}
In such systems with multiple repulsive length scales, a variety of cluster phases are possible due to the
combination of multi-scale repulsive interactions with attractive
ranges.~\cite{varney_hierarchical_2013} Some of the phases obtained in
Ref.~\onlinecite{varney_hierarchical_2013} were also reproduced in
simulations of a layered Ginzburg-Landau
model.~\cite{komendova_soft_2013} In
Ref.~\onlinecite{meng_honeycomb_2014}, it was discussed that these
models also allow vortex lattices with different symmetries. However a
phase diagram of these models has not been investigated previously.

In what follows, we will discuss the phenomenological model used in
this study in Sec.~\ref{sec:model}. In Sec.~\ref{sec:method}, the
details of the simulation method and measurements used to characterize
each phase are described. In Sec.~\ref{sec:results}, we present two
phase diagrams featuring 10 and 17 phases, respectively, and
characterize each phase. 
Discussion of the especially interesting phases in both phase diagrams
is in Sec.~\ref{sec:dis}.
Finally, the impact of the results are discussed in
Sec.~\ref{sec:summary}.

\section{\label{sec:model}Model}

 Motivated by the multi-scale character of intervortex interaction
  in type-1.5 systems, here we study a phase diagram of a
prototypical system with $N_v$ vortices exhibiting multi-scale
inter-vortex interactions. The effective model for a vortex under
these conditions can be described by using a hard-core radius
$\sigma_h$, an inner soft-core radius $\sigma_1$, and an outer
soft-core radius $\sigma_2$. In addition, we can account for the
effect of stray fields, which give an effective power-law
interaction\cite{pearl_current_1964}. To generate such potentials, we define the potential
energy using a phenomenological
model~\cite{varney_hierarchical_2013,meng_honeycomb_2014}
\begin{align}
  \frac{V(r)}{V_0} = e^{-r/\lambda} - c_2 e^{-r/\xi} + c_3
  \frac{\lambda \{\tanh[a(r - b)] + 1\}}{r +
    \delta}
  \label{eq:potentials}
\end{align}
where $V_0$ defines arbitrary unit of energy and $\lambda$ and $\xi$
play the role of penetration and the longest of the coherence lengths
(in the type-1.5 regime one of the coherence
lengths is the longest length scale and is associated with the
long-range intervortex attraction, yet vortices are thermodynamically
stable and form in an external
field\cite{babaev_semi-meissner_2005}). Since we are not interested in
very short range intervortex interaction we neglect detail of
interactions associated with shorter coherence lengths that are
present in Eq.~\eqref{eq:inn}. We set $\lambda=1$ to set the unit of
length and refer to all distances in terms of $\lambda$.
% = (\phi_0^2)/(2\pi\mu_0\lambda^3)$ defines the unit of the energy,
% $\phi_0$ is the flux quantum, $\mu_0$ is the vacuum permeability,
% $\lambda$ is the magnetic field penetration length, $\xi$ is the
% effective coherence length,
The constants $c_2, c_3, a, b$, and $\delta$ are phenomenological
coefficients. In order to get a potential qualitatively similar to
Eq.~\eqref{eq:inn} with the addition of a long-range interaction
caused by stray fields, we set $c_2=0.2$, $c_3=0.1$, and
$\delta=0.1$. In order to control the short-range repulsive length
scale of the inter-vortex potential, we tune $a$, $b$ and the ratio
$\xi/\lambda$.

The three terms in Eq.~\eqref{eq:potentials} are a short-range
repulsive interaction, an intermediate attractive interaction, and a
long-range repulsive interaction caused by stray
fields.~\cite{pearl_current_1964} In the denominator of the third
term, the parameter $\delta$ removes the hard core of the vortex,
i.e. $\sigma_h = 0$, and reduces the impact of the short-range
repulsive interaction caused by power-law term. For other contexts 
where multi-scale inter-particle interactions arise see Refs. 
~\onlinecite{malescio_stripe_2004, glaser_soft_2007,
  O.Reichhardt2010, O.Reichhardt2011, PhysRevE.85.051401, 
  malescio_stripe_2003, spivak_phases_2004, parameswaran_typology_2012}.

\begin{figure}[t]
  \includegraphics[width=\columnwidth]{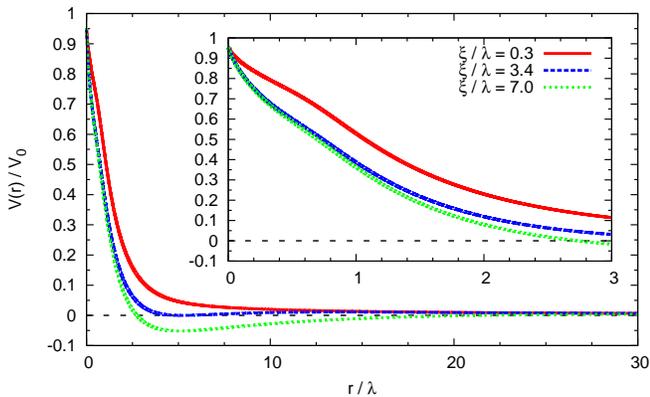}
  \caption{(Color online) Inter-vortex pair potentials at $a=2.5$,
    $b=0.5$ and $\xi/\lambda = 0.3$, $3.4$, and $7.0$. When $0.5 \leq
    \xi \leq 3.4$, there is only purely repulsive interactions in the
    potential. When $3.4 < \xi \leq 10.0$, the potential has the
    attractive well around $r/\lambda \approx 5.2$.
    \label{fig:a25b05allpotentials}
  }
\end{figure}

\begin{figure}[t]
  \includegraphics[width=\columnwidth]{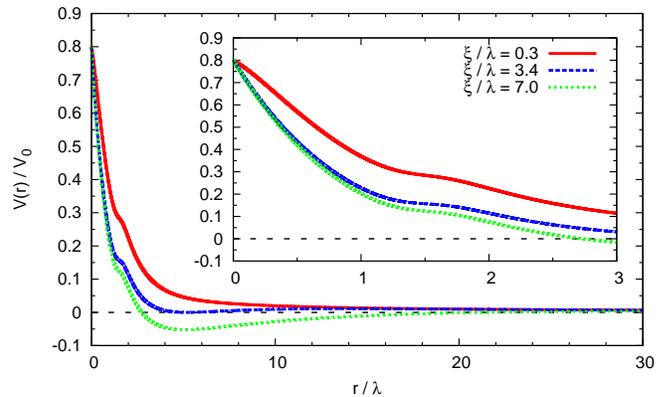}
  \caption{ (Color online) Inter-vortex pair potentials at $a=3.0$,
    $b=1.5$ and $\xi/\lambda = 0.3$, $3.4$, and $7.0$. Similarly as
    Fig.~\ref{fig:a25b05allpotentials}, when $\xi > 3.4$, there is one
    attractive well around $r/\lambda \approx 5.2$ at the potential.
    \label{fig:a30b15allpotentials}
  }
\end{figure}

In this work, we consider two series of potentials. In the first,
shown in Fig.~\ref{fig:a25b05allpotentials}, $a = 2.5$, $b = 0.5$, and
$\xi/\lambda$ ranges from $0.2$ to $10.0$. In the second, shown in
Fig.~\ref{fig:a30b15allpotentials}, $a = 3.0$, $b = 1.5$, and
$\xi/\lambda$ varies from $0.1$ to $10.0$. In both
Figs.~\ref{fig:a25b05allpotentials} and \ref{fig:a30b15allpotentials},
we show three typical potentials by changing the ratio of the length
scales $\xi/\lambda$. Note that as $\xi/\lambda$ increases, the
attractive region of the potentials are enhanced.

The solid red lines in Figs.~\ref{fig:a25b05allpotentials} and
\ref{fig:a30b15allpotentials} represent a typical potential form at
small $\xi$ values($\xi/\lambda = 0.3$), which contains three
short-range repulsive length scales and one long-range repulsive
length scale. The dashed blue lines in
Figs.~\ref{fig:a25b05allpotentials} and \ref{fig:a30b15allpotentials}
are for a potential at $\xi/\lambda = 3.4$, which has a minimum
$V(r)/V_0=0$ at $r/\lambda \approx 5.2$. From $\xi/\lambda = 3.4$ to
10.0, there is one attractive well in the potential and the attractive
short length increases as the $\xi$ increases. The dotted green line
in Figs.~\ref{fig:a25b05allpotentials} and
\ref{fig:a30b15allpotentials} represent a potential at
$\xi/\lambda=7.0$, which has an attractive well.

%There are two main differences between these two series of
%potentials. One is that the hard core radius $\sigma_h$ at
%$a=2.5,b=0.5$ is much smaller than $\sigma_h$ at $a=3.0,b=1.5$.  The
%other is that the first vortex potential shoulder height
%($\epsilon_2-\epsilon_1$) at $a=2.5,b=0.5$ is much larger than
%$a=3.0, b=1.5$.  Under this potential we can get the kagom\'e
%lattice\cite{QCHE}.

\section{\label{sec:method}Simulation method}
We utilize Langevin dynamics~\cite{fangohr_vortex_2001} with simulated
annealing to calculate the ground state of this vortex system. The
overdamped Langevin equation of motion for a vortex at $\textbf{r}_i$
is
\begin{align}
  {\bf F}_i = \sum_{j\ne i}^{N_v} {\bf F}^{vv}({\bf r}_i - {\bf r}_j)
  + {\bf F}_i^T = \eta \frac{d {\bf r}_i}{d t}
\end{align}
where ${\bf F}_i$ is the total force on vortex $i$, 
${\bf F}^{vv}({\bf r}_i - {\bf r}_j) = -\nabla V_{ij}$ is the
inter-vortex force between vortices $i$ and $j$, $N_v$ is the number
of vortices, ${\bf F}_i^T$ is the stochastic thermal force, $\eta$ is
the Bardeen-Stephen friction coefficient.

We do our simulation within a nearly square $L_x \times L_y$ box and
employ the periodic boundary conditions. We used random initial
configurations for each simulation and compared with a perfect
hexagonal lattice. In order to get the perfect initial hexagonal
lattice, $L_x$ and $L_y$ are chosen to alleviate the effects of
frustration\cite{fangohr_vortex_2001} and keep $L_x/L_y\approx1$. The
density is defined as $\rho=\frac{N_v}{L_x L_y}$, where $N_v$ is the
number of vortices. In the phase diagrams shown, the number of
vortices were $N_v = 780$ and $986$ at low and high density,
respectively. In addition, the existence of each phase was verified
with simulations of at least $N_v=2958$ vortices.

\begin{figure*}[t]
  \includegraphics[width=\textwidth]{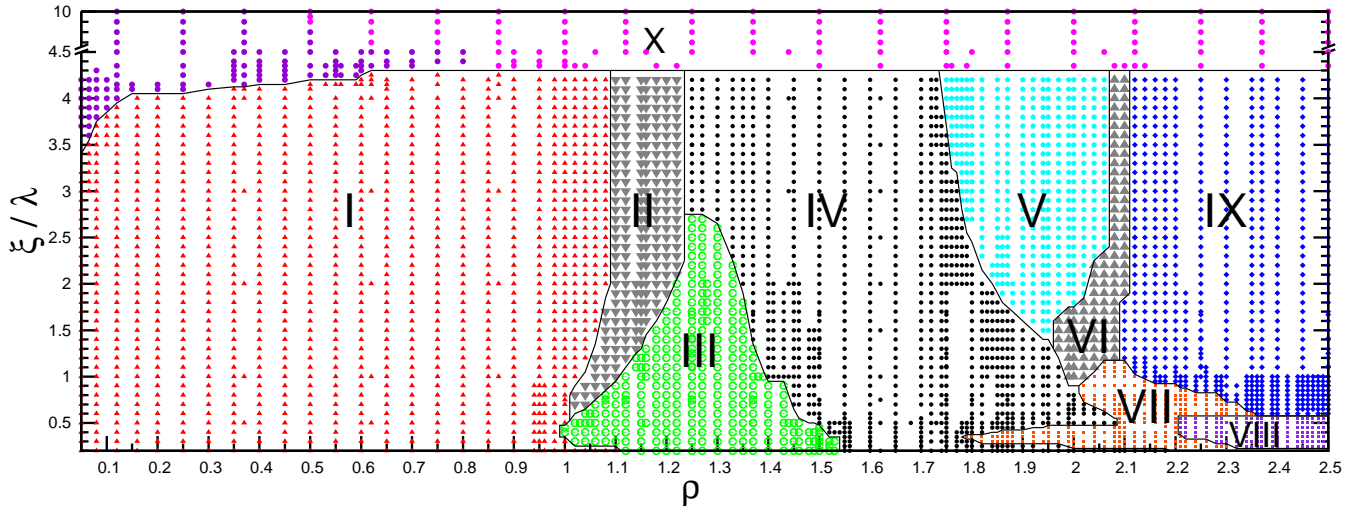}
  \caption{(Color online) Phase diagram of the final vortex
    configuration at zero temperature in the $\xi/\lambda$-$\rho$
    plane for the potential shown in
    Fig.~\ref{fig:a25b05allpotentials}. For each point, $N_v=780$ was
    used in the simulation with the exception of phases I and IV,
    which used $N_v=986$ and $N_v=4012$, respectively. Representative
    vortex configurations for each phase are shown in
    Fig.~\ref{fig:a25b05snapshots}.
    \label{fig:a25b05phases}
  }
\end{figure*}

\begin{figure*}[t]
  \includegraphics[width=\textwidth]{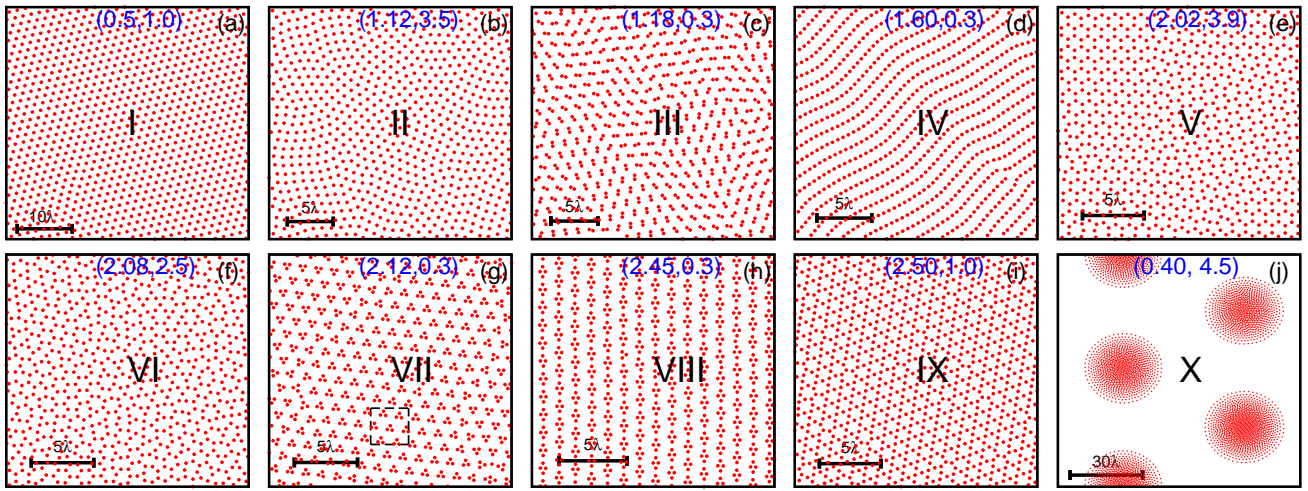}
  \caption{(Color online) Snapshot of a representative vortex
    configuration for each phase (indicated by Roman numerals) in
    Fig.~\ref{fig:a25b05phases}.  The coordinate of the phase diagram
    $(\rho,\xi/\lambda)$ is indicated on each panel. 
%    The corresponding RDFs and static structure factors are shown in
%    Figs.~\ref{fig:a25b05rdfs} and \ref{fig:a25b05sf}.
    \label{fig:a25b05snapshots}
  }
\end{figure*}

In order to characterize each phase, we first considered the radial
distribution function~(RDF)
\begin{align}
  g(r) = \frac{1}{2\pi r \Delta r \rho N_v} \sum_{i = 1}^{N_v} n_i(r,\Delta r),
\end{align}
where $n_i(r,\Delta r)$ is the number of particles in the shell
surrounding the $i$th vortex with radius $r$ and thickness $\Delta
r$. The RDF is used to characterize the structure of a
configuration. For small separations, $g(r)=0$ and as $r\to\infty$
$g(r)\to1$. In the RDF, sharp peaks are a clear signature of ordering
into shells of nearest neighbors while peak broadening is an
indication of a breakdown in long-range order. An absence of
well-defined peaks is a clear signature of local structure only.
% In RDF, if there are no sharp peaks, only broad peak and no
% explicitly minimum between different peaks, it indicates that there
% is no clearly local structure in this phase.

Next, we considered the static structure
factor~\cite{imperio_microphase_2006}
\begin{align}
  S({\bf k}) &= \frac{1}{N}\left\langle \left(\sum_{i}^N \cos({\bf k}
  \cdot {\bf r}_i)\right)^2 + \left(\sum_{i}^N \sin({\bf k} \cdot
  {\bf r}_i)\right)^2 \right\rangle
\end{align} 
to characterize the intermediate and large scale structure of the
vortex configurations.  Here $\mathbf{k} \equiv (k_x, k_y)$, $k_{x(y)}
= n_{x(y)} k_{0x(0y)}$ and $k_{0x(0y)}=2\pi/L_{x(y)}$. In our
simulations, $L_x \approx L_y$, so $k_{0x} \approx k_{0y}$ and we use
$k_{0x}$ as the unit of scale.

The final measures we use to define the structure is the number of
nearest neighbors $n_1$ and the nearest-neighbor distance $r_1$. The
number of nearest neighbors $n_1$ is determined by counting the
average number of vortices within circle of radius $r_\textrm{min}$
where $r_\textrm{min}$ is the distance corresponding to the minimum
value between the first and second peak in the RDF. The nearest
neighbor distance $r_1$ is defined by the location of the first
maximum in $g(r)$.

While ordered phases will have clear signatures in $g(r)$, $S({\bf
  k})$, and $n_1$, disordered phases will not exhibit long-range order
and are characterized by a uniform ring structure in $S({\bf k})$ and
lack of symmetry breaking. Phases that have these characteristics in
the structure factor will be referred to as as disordered or ``glassy" phase
(we do not discuss dynamic characterisation of glassiness in this paper).
 
\section{\label{sec:results}Results}
\subsection{Phase Diagram I}
% with interaction in Fig.~\ref{fig:a25b05allpotentials}}
The ground state phase diagram in the density $\rho$ % \cite{noterho} 
and  $\xi/\lambda$ plane corresponding to
Fig.~\ref{fig:a25b05allpotentials} is shown in
Fig.~\ref{fig:a25b05phases}. Here, we set $a = 2.5$, $b = 0.5$. The
density is varied from $\rho = 0.05$ to $2.50$ and $\xi/\lambda$
ranges from $0.2$ to $10.0$. The phase diagram exhibits 10 different
phases and representatives of each phase are illustrated in
Fig.~\ref{fig:a25b05snapshots}, where each phase is represented by a
different symbol. Phase I is a hexagonal lattice (solid red
up-pointing triangle), Phase III is a dimer lattice (green circle),
and Phase IV is a stripe phase (solid black circle). Phase V is a
honeycomb lattice (solid cyan pentagon), Phase VII is a polarized
triangular trimer (solid orange square), Phase VIII is a tetramer
lattice (solid purple square), and Phase IV is a kagom\'e
lattice (solid blue rhombus). Phase X is a cluster phase featuring
multiple clusters (solid violet circle) for smaller densities and
$\xi/\lambda$ and single giant clusters (solid pink circle) for larger
densities and $\xi/\lambda$. Phases II and VI are disordered states (solid
dark gray down-pointing and up-pointing triangles,
respectively). Phase II separates the hexagonal lattice and stripe
phases while Phase VI is bordered by the honeycomb and kagome lattice
phases. Consequently, we shall distinguish these two states as GHeS
and GHoK, which is an combination of the phases that border the disoredered
phases.

\begin{figure}[t]
  \includegraphics[height=0.68\textwidth]{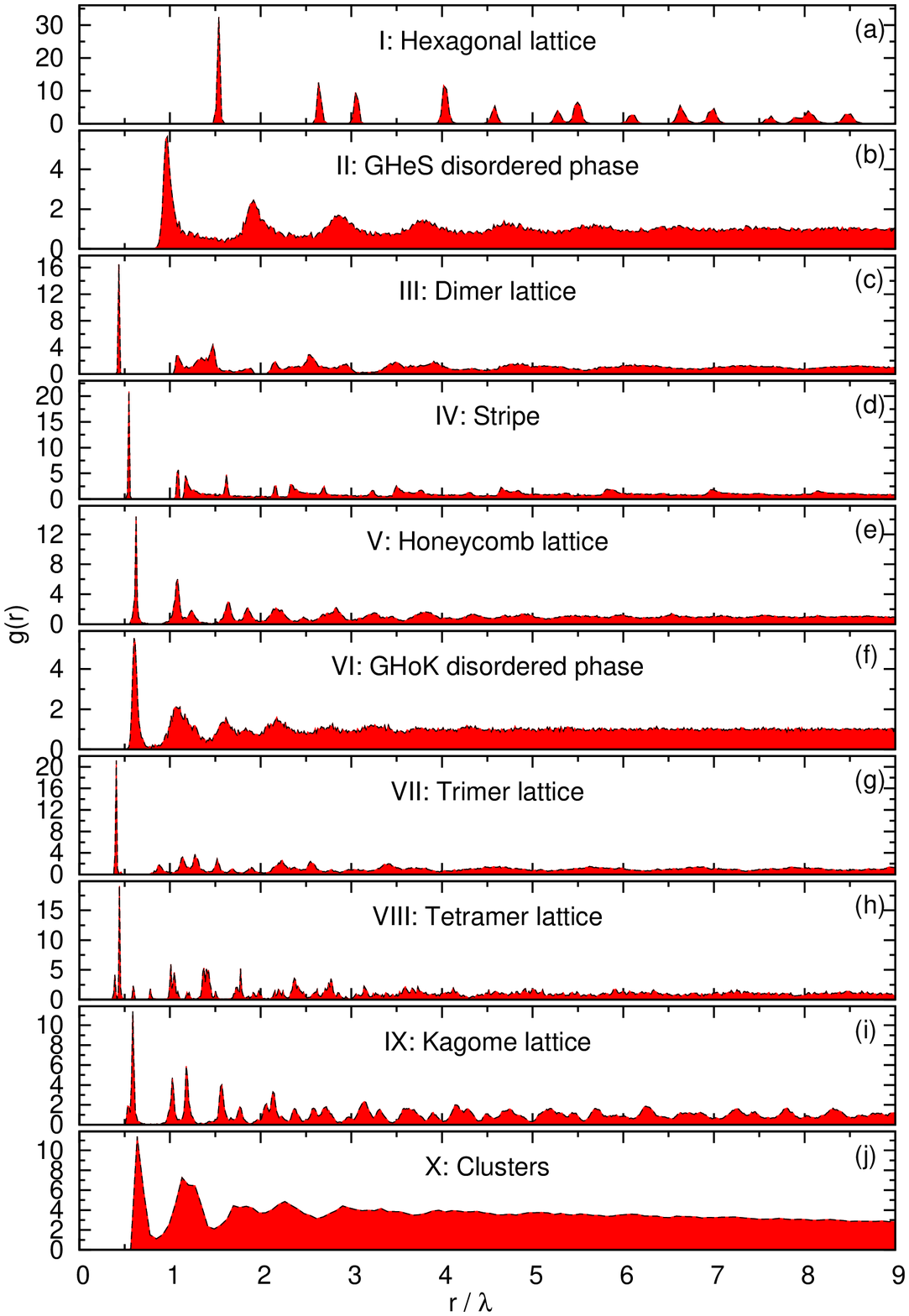}
  \caption{ (Color online) Radial distribution functions $g(r)$
    corresponding to the phases illustrated in
    Fig.~\ref{fig:a25b05snapshots}.
    \label{fig:a25b05rdfs}
  }
\end{figure}

\begin{figure*}[t]
  \includegraphics[width=\textwidth]{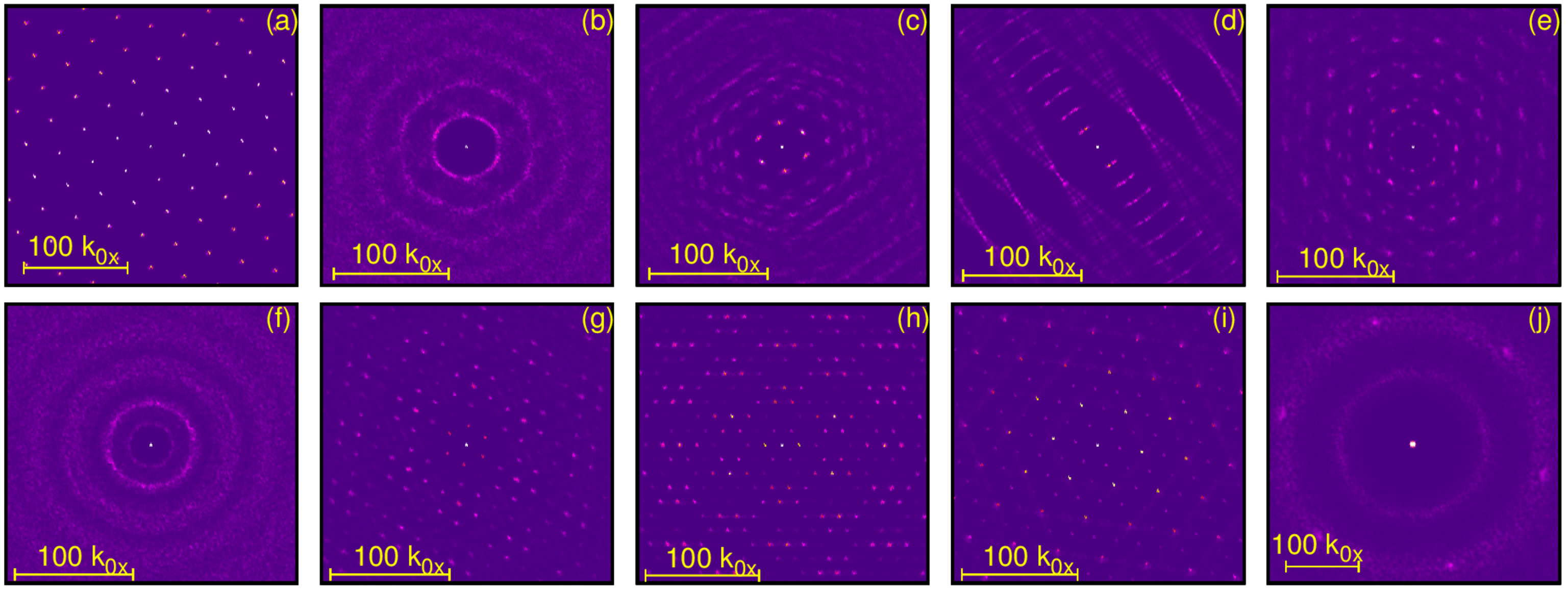}
  \caption{(Color online) Static structure factor $S({\bf k})$
    corresponding to the phases illustrated in
    Fig.~\ref{fig:a25b05snapshots}.
    \label{fig:a25b05sf}
  }
\end{figure*}

\begin{figure}[htb]
  \includegraphics[height=0.68\textwidth]{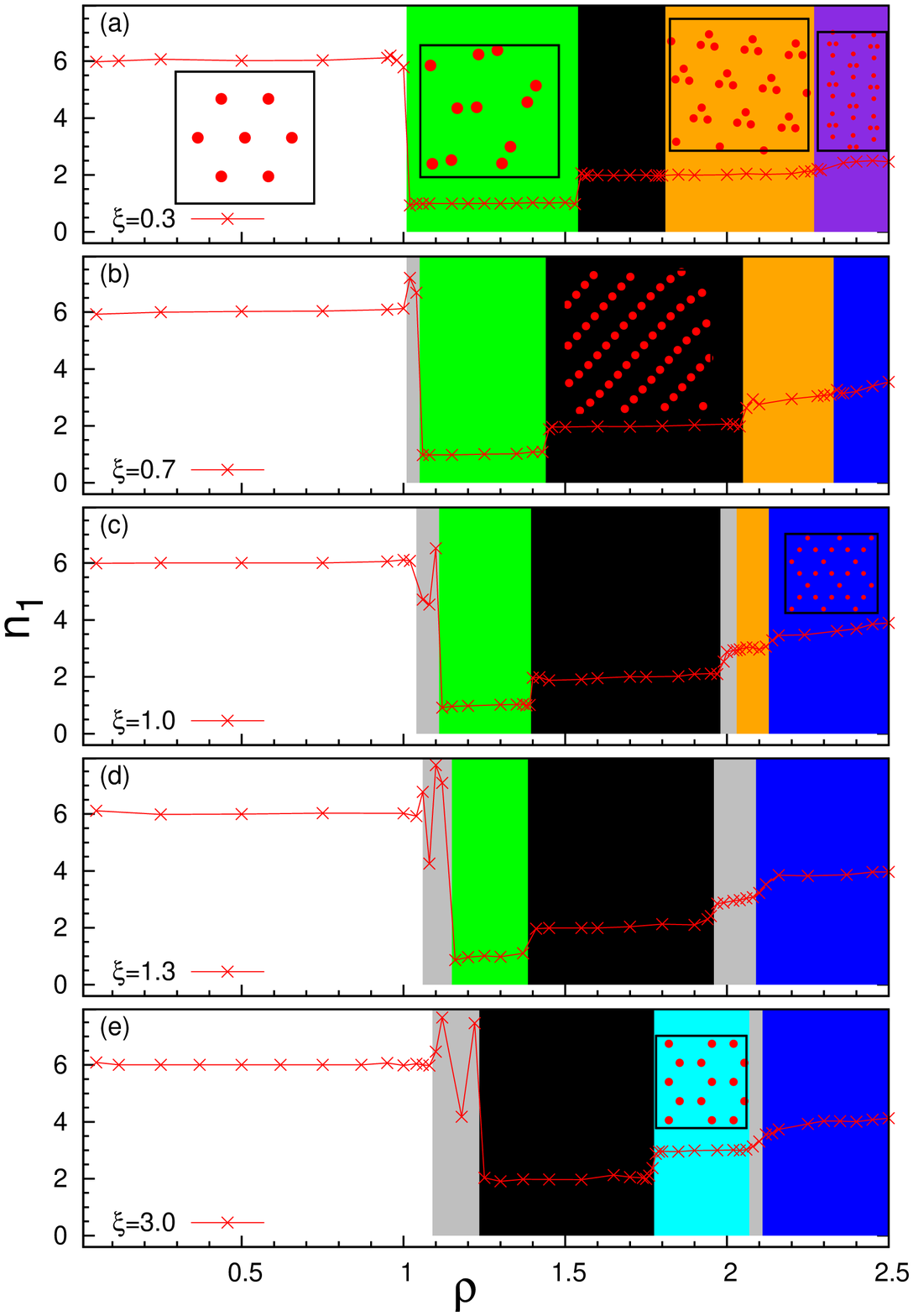}
  \caption{ (Color online) Nearest neighbor number $n_1$ at:
    (a). $\xi=0.3$; (b). $\xi=0.7$; (c). $\xi=1.0$; (d). $\xi=1.3$;
    (e). $\xi=3.0$. The background color corresponds to different
    phases: white (hexagonal), gray (disordered), green (dimer lattice),
    black (stripe), cyan (honeycomb lattice), orange (trimer lattice),
    purple (tetramer lattice), and blue (kagome lattice). There are
    two separate disordered ``glassy" phases, GHeS and GHoK, at $\rho \approx 1.00$
    and $\rho \approx 2.00$, respectively. The insets show typical
    configurations for each phase.
    \label{fig:n1alla25}
  }
\end{figure}

\begin{figure}[htb]
  \includegraphics[height=0.68\textwidth]{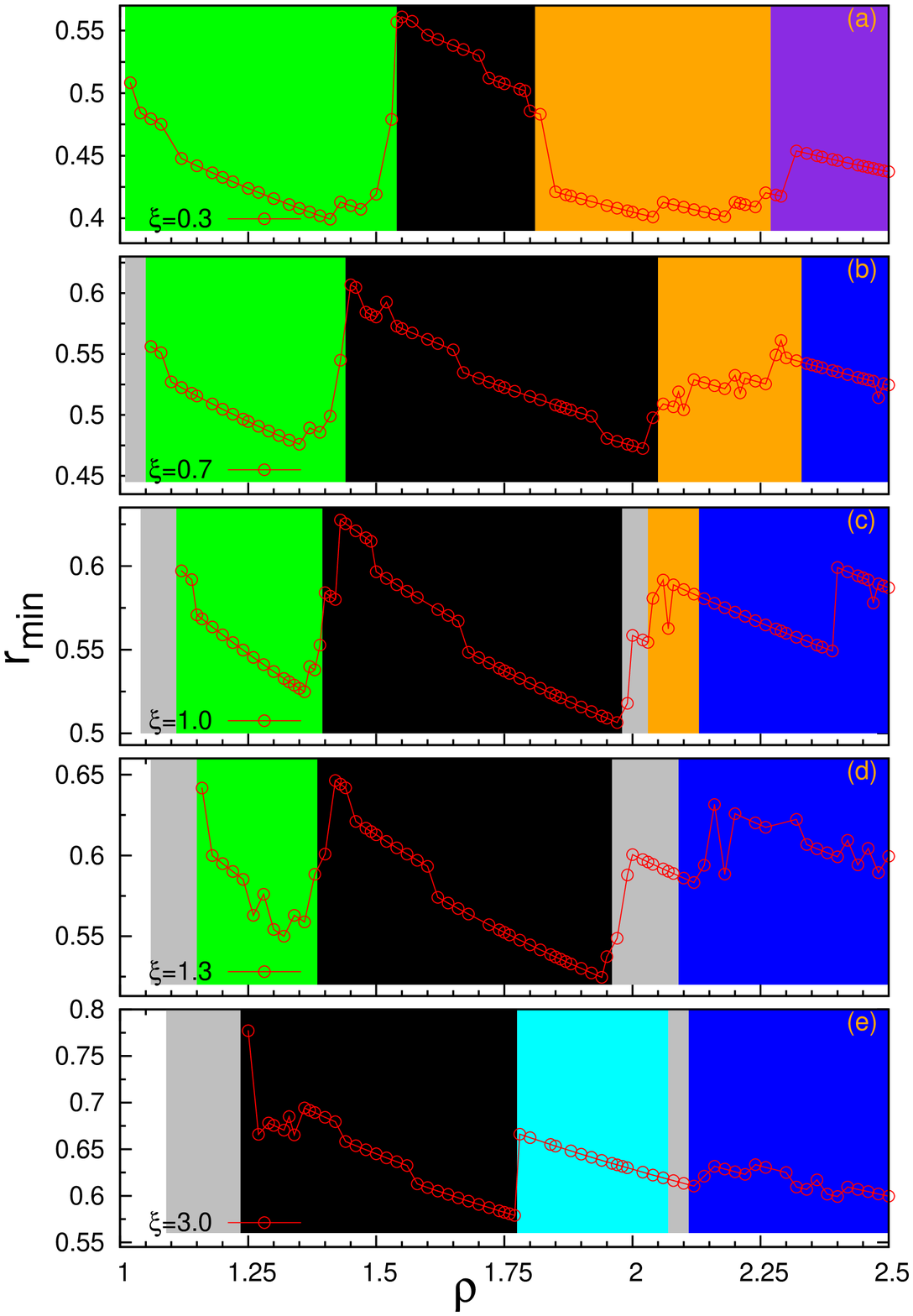}
  \caption{ (Color online) The distance between the nearest neighbor
    of the vortex $r_1$ as a function of density $\rho$. The
    background colors correspond to the same phases as described in
    Fig.~\ref{fig:n1alla25} and $r_1$ was determined by the position
    of the first peak of $g(r)$.
    \label{fig:rminalla25}
  }
\end{figure}

Figure~\ref{fig:a25b05snapshots} shows a snapshot of a representative
state for every phase in Fig.~\ref{fig:a25b05phases}. The
corresponding radial distribution functions $g(r)$ and static
structure factors $S(\mathbf{k})$ are shown in
Figs.~\ref{fig:a25b05rdfs} and \ref{fig:a25b05sf}, respectively. The
number of the nearest neighbors $n_1$ in each phase and the nearest
neighbor distance are shown in Fig.~\ref{fig:n1alla25} and
Fig.~\ref{fig:rminalla25}, respectively, for various $\xi/\lambda$ to
show how each quantity changes along all of the phase transitions
Fig.~\ref{fig:a25b05phases}. In Figs.~\ref{fig:n1alla25} and
\ref{fig:rminalla25}, the background colors correspond to different
phases: white (hexagonal lattice), gray (disordered ``glassy" phases), green (dimer
lattice), black (stripe), cyan (honeycomb lattice), orange (trimer
lattice), purple (tetramer lattice) and blue (kagome
lattice). Note that the gray regions corresponding to both disordered
phases, with the GHeS phase at $\rho \approx 1.00$ and GHoK at $\rho
\approx 2.00$. Here we show the $n_1$ for $\xi=0.3$,
$0.7$, $1.0$, $1.3$, and $3.0$, which is sufficient to show all of the phase
transitions in Fig.~\ref{fig:a25b05phases}. 

Let us first consider the phases that exhibit long-range
order. For $\xi/\lambda<3.5$ and $\rho=0.05$ to $\rho\approx1.00$, the
ground state of the system is a hexagonal lattice (Phase I in
Fig.~\ref{fig:a25b05phases}). A typical ground state configuration is
shown in Fig.~\ref{fig:a25b05snapshots}(a) and corresponds to
$\rho=0.50$ and $\xi/\lambda=1.0$. This phase is characterized by
sharp peaks in $g(r)$ located at distances corresponding to the
neighbor distances of a hexagonal lattice and a sharply defined
hexagonal structure in $S({\bf k})$, which are shown in
Fig.~\ref{fig:a25b05rdfs}(a) and Fig.~\ref{fig:a25b05sf}(a),
respectively. Another confirmation of the structure can be found in
Fig.~\ref{fig:n1alla25}, where the hexagonal lattice is indicated by a
white background and has $n_1=6$. The prevalence of this phase at low
densities is largely driven by the long-range repulsive interaction
caused by stray-fields.

\begin{figure}[ht]
  \includegraphics[height=0.18\textwidth]{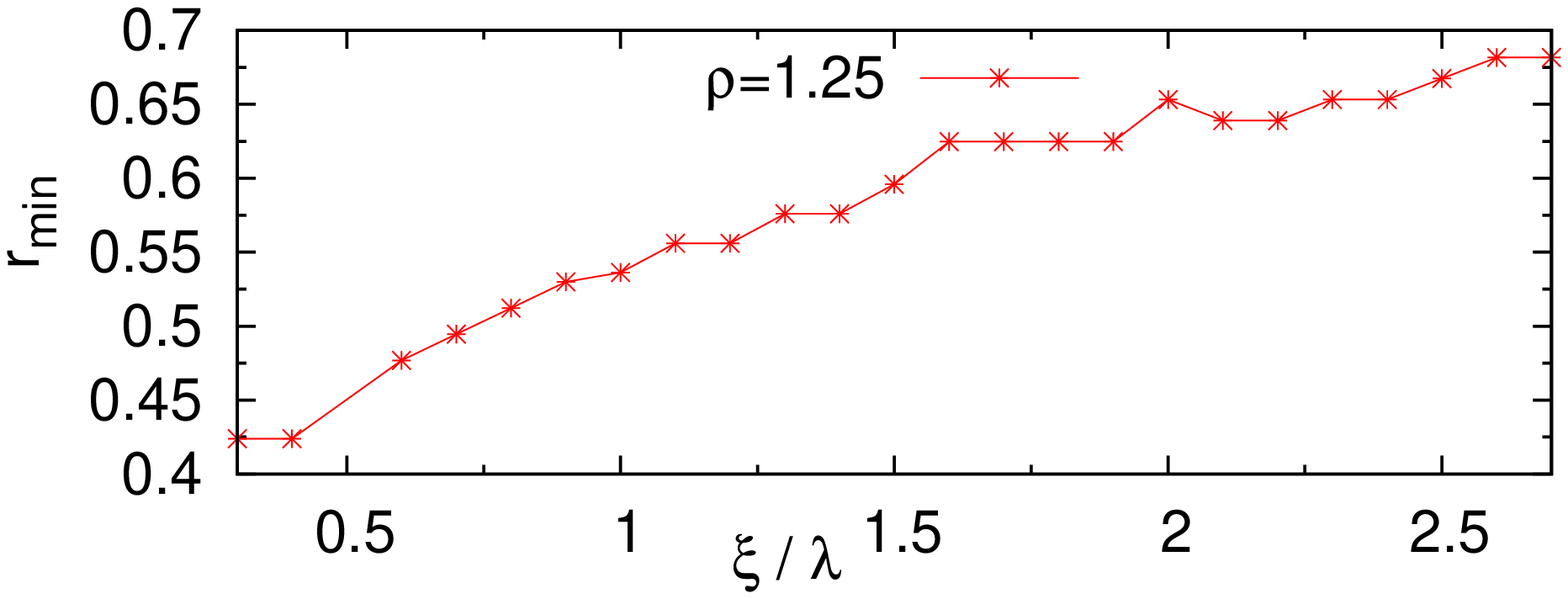}
  \caption{ (Color online) The distance between the nearest neighbor
    of the vortex $r_1$ as a function of  
    $\xi/\lambda=0.3$ to 2.7 at fixed $\rho=1.25$.
    \label{fig:rminrho125a25}
  }
\end{figure}

\begin{figure}[t]
  \includegraphics[height=0.23\textwidth]{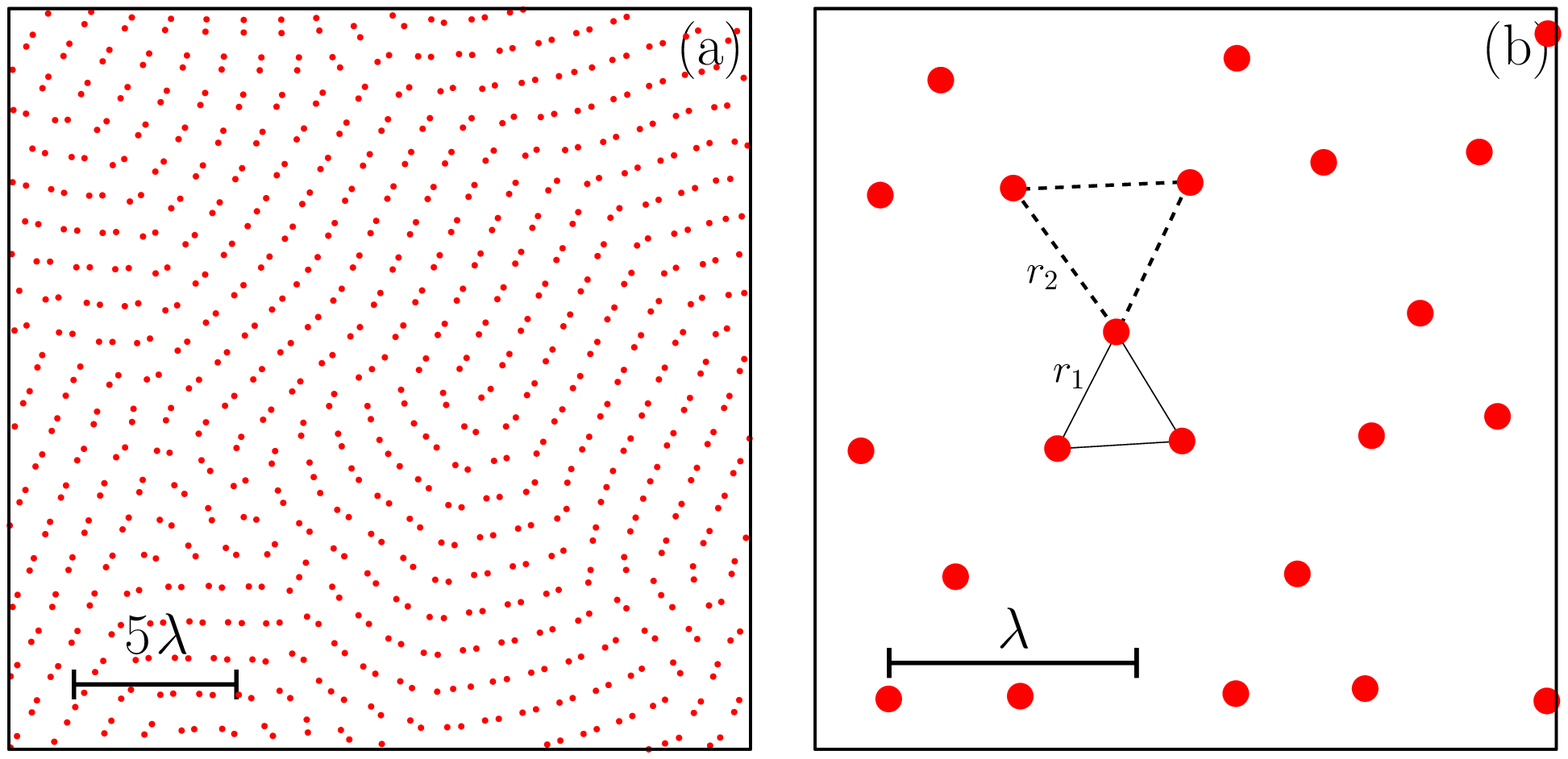}
  \caption{ (Color online) (a) Dimer lattice lined up at $\rho=1.50$
    with $\xi/\lambda=0.3$. (b) A small fraction of trimer lattice at
    $\rho=2.26$, $\xi/\lambda=0.8$, $N_v=780$. $r_1$ is the distance
    between vortices within one trimer and $r_2$ is the distance
    between different trimers. $r_1$ and $r_2$ is used to explain the
    transformation from trimer lattice to kagome lattice.
    \label{fig:a25b05dimerrho150}
  }
\end{figure}

Next, for small $\xi/\lambda$ and densities ranging from $\rho=1.0$ to
$\rho=1.5$, vortices continue to form a hexagonal lattice but
individual vortices pair to form dimers (Phase III). A typical
snapshot of a dimer lattice ($\rho=1.18$ and $\xi/\lambda=0.3$) is
shown in Fig.~\ref{fig:a25b05snapshots}(c), where there is some
alignment of the dimers but domains form for smaller $\rho$ and
$\xi/\lambda$. This phase is similar to the dimer phase observed in
potentials with two repulsive length
scales~\cite{malescio_stripe_2004,glaser_soft_2007,
  O.Reichhardt2010,O.Reichhardt2011}. In $g(r)$, see
Fig.~\ref{fig:a25b05rdfs}(c), the location of the first peak is at
$r/\lambda=0.44$ while the first peak of the hexagonal lattice is at
$r/\lambda=0.99$~\cite{note1}. Moreover, Fig.~\ref{fig:n1alla25}
clearly indicates that a sharp transition as $n_1\to1$ upon entering
the dimer lattice phase. The static structure factor shown in
Fig.~\ref{fig:a25b05sf}(c) displays a six-fold rotational
symmetry. However, at large ${\bf k}$ the intensity of the peaks is
diminished and the peaks are broadened due to the ordering of the
dimers into domains of alignment. As the density and $\xi$ are increased, the dimers in Phase III begin to align.  At
fixed $\xi/\lambda$, increasing the density results in a decrease in
$r_1$ (see. Fig.~\ref{fig:rminalla25}) while increasing $\xi/\lambda$
at fixed $\rho$ results in an increases in $r_1$ (see
Fig.~\ref{fig:rminrho125a25}). An example configuration of aligned
dimers is shown in Fig.~\ref{fig:a25b05dimerrho150}(a).

Further increasing the density and/or $\xi/\lambda$ results in
the formation of stripes in the ground state (Phase IV). A
representative configuration is shown in
Fig.~\ref{fig:a25b05snapshots}(d) and corresponding $g(r)$ and $S{(\bf
  k})$ are plotted in Fig.~\ref{fig:a25b05rdfs}(d) and
Fig.~\ref{fig:a25b05sf}(d), respectively. From the RDF for $\rho=1.60$
and $\xi/\lambda=0.3$, we observe the first peak at $r_1=0.546$ and
subsequent peaks at $r_2=1.093$, $r_3=1.171$, and $r_4=1.624$. The
second and fourth peaks are correlated with the line of vortices in a
stripe with $r_2=2 r_{1}$ and $r_4\approx3r_1$. The third peak,
however, describes the distance between neighboring stripes. In the
structure factor, we observe a dispersed two-fold symmetry
corresponding to the stripe directions. The distance between nearest
neighbor $r_1$ will also decrease as the density increases which was
shown in the black section of Fig.~\ref{fig:rminalla25}.

\begin{figure}[ht]
  \includegraphics[height=0.36\textwidth]{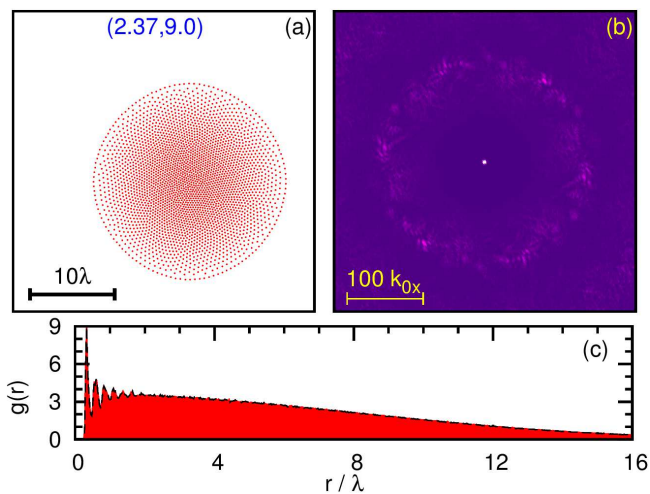}
  \caption{ (Color online) The (a) vortex configuration, (b) static
    structure factor, and (c) RDF of a single cluster with
    $\rho=2.37$, $\xi/\lambda=9.0$, and $N_v=2958$.
    \label{fig:a25clusterCombine}
  }
\end{figure}

\begin{figure}[ht]
  \includegraphics[height=0.36\textwidth]{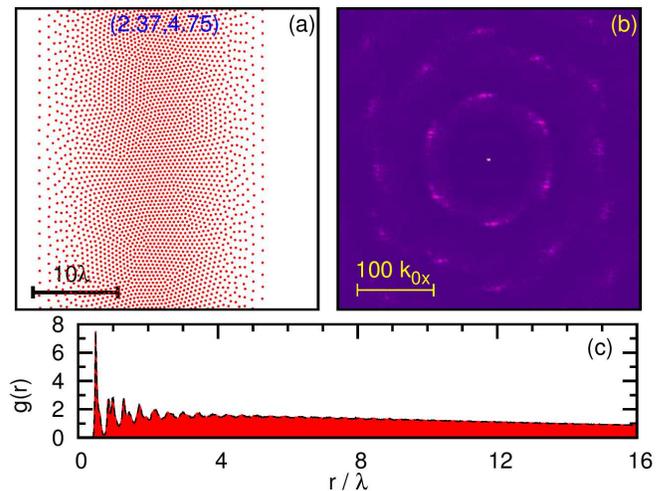}
  \caption{ (Color online) The (a) vortex configuration, (b) static
    structure factor and (c) RDF of the non-uniform stripe phase at
    $\rho=2.37$, $\xi/\lambda=4.75$, and $N_v=2958$. When the vortices
    number increases to $N_v=4012$, the whole system will form a
    circular cluster.
    \label{fig:a25clusterMiddle}
  }
\end{figure}

\begin{figure}[ht]
  \includegraphics[height=0.36\textwidth]{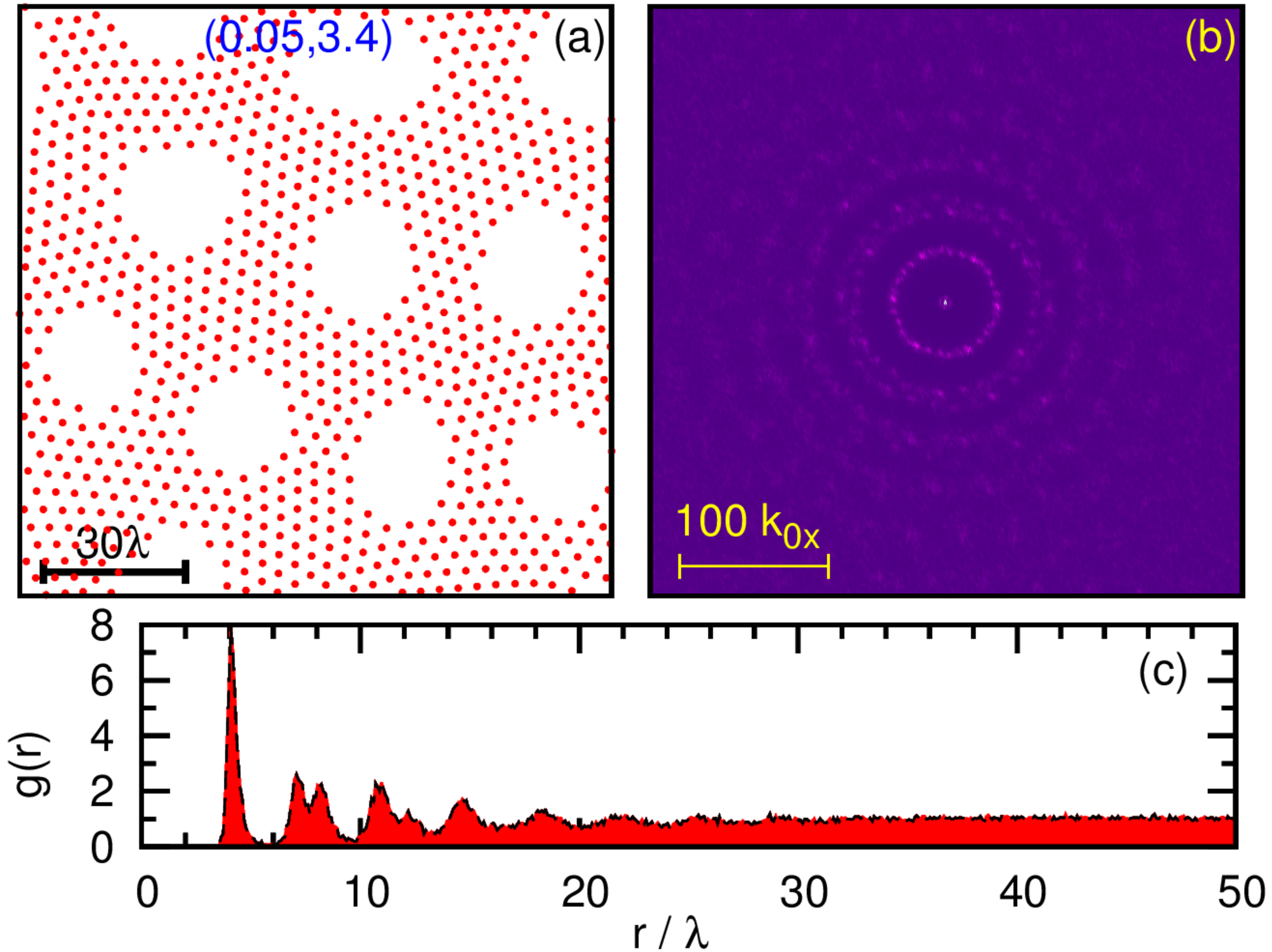}
  \caption{ (Color online) The (a) vortex configuration, (b) static
    structure factor, and (c) RDF of the void phase at $\rho=0.05$,
    $\xi/\lambda=3.4$, and $N_v=780$.
    \label{fig:a25clusterVoid}
  }
\end{figure}

As the density $\rho$ continues to grow, the stripe phase will
transform to honeycomb lattice for $\xi/\lambda > 1.5$, a disordered phase
for $1.0<\xi/\lambda<1.5$, and a trimer lattice for
$\xi/\lambda<1.0$. The representative snapshot ($\rho=2.02$ and
$\xi/\lambda=3.9$) for the honeycomb lattice (Phase V) is shown in
Fig.~\ref{fig:a25b05snapshots}(e). The radial distribution function
[Fig.~\ref{fig:a25b05rdfs}(e)], the structure factor
[Fig.~\ref{fig:a25b05sf}(e)], and nearest neighbor number $n_1=3$ are
consistent with what is expected for a honeycomb lattice but the phase
contains a number of defects that result in a broadening in the peaks
of $g(r)$ and $S({\bf k})$.

From $0.2 \leq \xi < 1.0 $, the stripe will transform to a polarized
triangle trimer lattice, which was denoted as Phase VII in the phase
diagram in Fig.~\ref{fig:a25b05phases}. (The reason why it's called
polarized triangle trimer lattice will be discussed later in
Fig.~\ref{fig:trimerCompare}.) A representative snapshot and the
corresponding RDF and static structure factor $S(\mathbf{k})$ for this
phase at $\rho=2.12$, $\xi/\lambda=0.3$ are shown in
Fig.~\ref{fig:a25b05snapshots}(g), \ref{fig:a25b05rdfs}(g) and
\ref{fig:a25b05sf}(g), respectively. Similar to a stripe phase, one
vortex in a trimer lattice has two nearest neighbors located at a
distance $r_1$ and form an equilateral triangle. Each trimer then
forms a hexagonal lattice. Consequently, the RDF has a sharp first
peak and subsequent peaks correspond to the distance between vortices
in different trimers. The static structure factor $S(\mathbf{k})$
pattern shown in Fig.~\ref{fig:a25b05sf}(g) shows a hexagonal lattice
with six voids around one hexagon which is same as the result of
perfect trimer lattice created by artificially pinning the
vortices.~\cite{PhysRevE.85.051401}
%%%%%%%%%%%%%%%%%%%%%%%%%%%%%%%%%%%%%%%%%%%%%%%%%%%%%%%%%%%%%%%%%%%%
% recovered since I think it's important to explain why the $n_{1}$
% changes in Fig.~\ref{fig:n1alla25}(a).
%%%%%%%%%%%%%%%%%%%%%%%%%%%%%%%%%%%%%%%%%%%%%%%%%%%%%%%%%%%%%%%%%%%%
The micro-structure of the polarized triangle trimer lattice is shown
in the pinned panel in the orange section in
Fig.~\ref{fig:n1alla25}(a). The nearest neighbor number $n_1$ changes
at different $\xi/\lambda$. For $0.2 \leq \xi < 0.6$, $n_1$ will
always be 2 which is shown in the orange section in
Fig.~\ref{fig:n1alla25}(a). For $0.6 \leq \xi < 1.2$, $n_1$ will
increase from 2 to 3 which is shown in the orange section in
Fig.~\ref{fig:n1alla25}(b,c). The reason will be discussed later at
the transformation from trimer lattice to kagome lattice. The distance
between nearest neighbor $r_1$ is not always decrease as the density
$\rho$ increases which is shown in the orange section in
Fig.~\ref{fig:rminalla25}(a,b,c).
%%%%%%%%%%%%%%%%%%%%%%%%%%%%%%%%%%%%%%%%%%%%%%%%%%%%%%%%%%%%%%%%%%%%%% 

As the density continues to grow, the trimer lattice at
$0.2\leq\xi<0.6$ transforms to a tetramer lattice (Phase VIII). The
representative snapshot, RDF, and static structure factor
$S(\mathbf{k})$ for this phase at $\rho=2.45$ and $\xi/\lambda=0.3$
are shown in Figs.~\ref{fig:a25b05snapshots}(h),
\ref{fig:a25b05rdfs}(h) and \ref{fig:a25b05sf}(h), respectively. In
this phase, four vortices will form a rhombus group due to the
short-range interaction and the long-range interaction forces the
tetramers to order into a hexagonal lattice. In a tetramer, half the
vortices have 2 nearest neighbors and half have 3 nearest neighbors,
resulting in $n_1 = 2.5$, which can be clearly seen in
Fig.~\ref{fig:n1alla25}(a). The micro-structure of the tetramer
lattice is shown in the pinned panel in the purple section in
Fig.~\ref{fig:n1alla25}(a). The distance between nearest neighbor
$r_1$ decreases as the density increases which is shown in the purple
section in Fig.~\ref{fig:rminalla25}(a).

For $0.6 \leq \xi < 1.2$ the trimer lattice transforms to a kagome
lattice (Phase IX). The representative snapshot, RDF, and static
structure factor for the kagome lattice at $\rho=2.50,
\xi/\lambda=1.0$ are shown in Fig.~\ref{fig:a25b05snapshots}(i),
\ref{fig:a25b05rdfs}(i), and \ref{fig:a25b05sf}(i), respectively. In
this phase each vortex has 4 nearest neighbors and $g(r)$ features a
sharp peak corresponding to the nearest neighbor distance, with
subsequent peaks being related to the distances between the second,
third, and fourth nearest neighbors.

The transition between the trimer and kagome lattice phases can be
understood more clearly by carefully examining the trimer lattice
phase. In Fig.~\ref{fig:a25b05dimerrho150}(b), we illustrate an
individual trimer and its neighbors at $\rho=2.26$, $\xi=0.8$,
$N_v=780$. Here, $r_1$ is the distance between vortices within one
trimer and $r_2$ is the distance between neighboring trimers. As the
density increases, the difference between $r_1$ and $r_2$ continues to
decrease. Once $r_1=r_2$, the system forms a kagome lattice since each
trimer has the same alignment.

The transition from the honeycomb lattice (Phase V) to the kagome
lattice (Phase IX) is separated by a disordered region with
characteristics of both the honeycomb and kagome lattices, which will
be referred to as GHoK (Phase VI). A representative snapshot of this
phase at %$\rho=2.07$ and $\xi/\lambda = 3.0$ (or 
$\rho=2.08$ and $\xi/\lambda=2.5$ is shown in
Fig.~\ref{fig:a25b05snapshots}(f). Here, the system forms domains of
honeycomb and kagome lattices with the size of the kagome domains
increasing as the density increases and the number of nearest
neighbors smoothly increasing from $n_1=3$ to $n_1=4$ (see
Fig.~\ref{fig:n1alla25}). The RDF and static structure factor
corresponding to this snapshot are illustrated in
Figs.~\ref{fig:a25b05rdfs}(f) and \ref{fig:a25b05sf}(f). There are no
sharp peaks in $g(r)$, shown in Fig.~\ref{fig:a25b05rdfs}(f), and
there are only broad peaks and plateaus with no clear minima. The
static structure factor in Fig.~\ref{fig:a25b05sf}(f) shows four
uniform rings which suggests that this phase has no broken symmetries
and is a disordered  phase.

% In Fig.~\ref{fig:n1alla25}(a), from $\rho=1.01$ to 2.50, $n_1$ will
% change from 1 to 2 and finally to 2.5. In
% Fig.~\ref{fig:n1alla25}(b,c,d), from $\rho=1.15$ to 2.50, $n_1$ will
% change from 1 to 2 and finally to 4. In Fig.~\ref{fig:n1alla25}(e),
% from $\rho=1.235$ to 2.50, $n_1$ will change from 2 to 3 and finally
% to 4.

In addition to the disordered phase separating the honeycomb and kagome
lattices, there exists a disordered phase separating the hexagonal lattice
(Phase I), dimer lattice (Phase III), and stripe (Phase IV) phases,
which we will refer to as GHeS (Phase II). The transition occurs
directly from the hexagonal lattice for $\xi > 0.6$. As the density
increases further, the disordered phase transitions to the dimer lattice
for $\xi < 2.75$ and the stripe phase for $\xi > 2.75$. A
representative snapshot of the GHeS phase at $\rho=1.12$,
$\xi/\lambda=3.5$ is shown in Fig.~\ref{fig:a25b05snapshots}(b) and
the corresponding $g(r)$ and $S(\mathbf{k})$ are shown in
Fig.~\ref{fig:a25b05rdfs}(b) and \ref{fig:a25b05sf}(b),
respectively. From the $g(r)$, we can see that the position of the
first peak of GHeS phases, $r/\lambda=0.97$, is closer than the
position of the first peak in hexagonal lattice,
$r/\lambda=1.02$.~\cite{note1} Note that the peaks in $g(r)$ are much
broader than the hexagonal lattice result and that there are plateaus
and no single clear minimum between peaks. All of these features
suggest that it has no clear local structure. In the static structure
factor, there are several rings and no symmetry is broken. The
combination of the features in the RDF and the static structure factor
indicate that Phase II is a frustrated disordered phase. From the
representative snapshots in Fig.~\ref{fig:a25b05snapshots}(b), we note
that the hexagonal order has been destroyed and that the system is
attempting to form a linear order but there is too much competition
for it to occur. In this phase, the number of nearest neighbors varies
wildly, either increasing to around 7 or decreasing to around 4 (see
Fig.~\ref{fig:n1alla25}).

\begin{figure*}[!t]
  \includegraphics[width=1.0\textwidth]{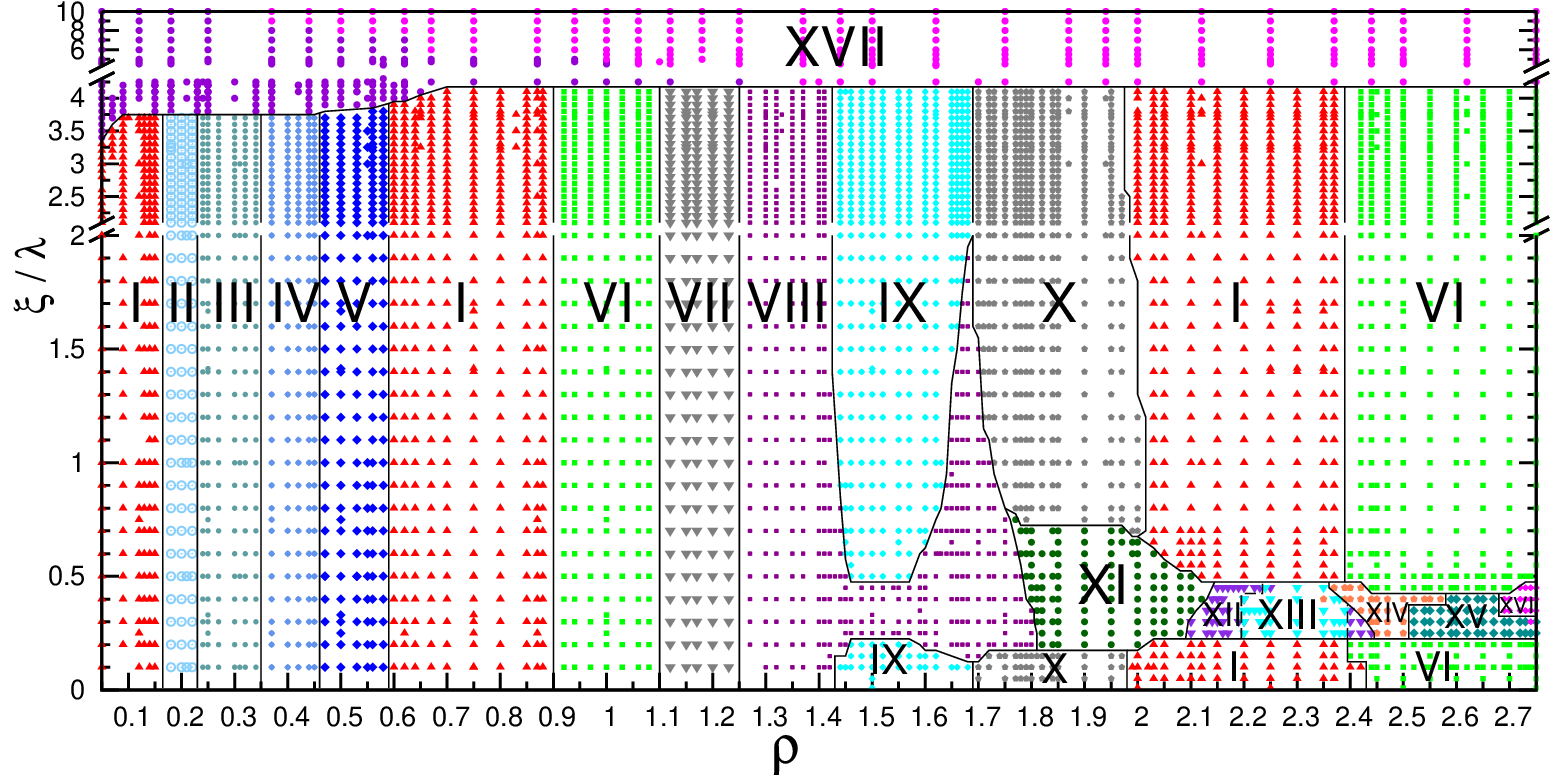}
  \caption{ (Color online) Phase diagram of the final vortex
    configuration at zero temperature in the  
    $\xi/\lambda$ and density $\rho$ plane for the potentials shown in
    Fig.~\ref{fig:a30b15allpotentials}. There are seventeen phases in
    this phase diagram. The Roman numerals in every section denote the
    every phase. The snapshot, RDF, and static structure factor of a
    representative final configuration for each phases is shown in
    Figs.~\ref{fig:a30b15snapshots}, \ref{fig:a30b15rdfs}, and
    \ref{fig:a30b15sf}, respectively.
    \label{fig:a30b15phases}
  }
\end{figure*}

 Generally, in the phase diagram in Fig.~\ref{fig:a25b05phases}, the
long range power law interaction will force the vortices to form a
hexagonal lattice at low densities. As the density increases, the
short-range repulsive and attractive interactions will dominate with
the long-range term, forcing the system to form different phases as
the short-range repulsive and attractive terms compete. But there
exists a  rule: under the pair interaction in
Fig.~\ref{fig:a25b05allpotentials}, the short range interaction will
force the system to form a higher nearest neighbor number $n_1$ phase
as the density $\rho$ increases. Moreover, the presence of the
long-range power law term enhances long-range ordering of the
local patterns.  When the density increases significantly, the
short-range repulsive term becomes especially important resulting  in a
frustrated kagome lattice.

However, when the attractive term is enhanced by increasing $\xi$
above a critical $\xi_c$, it dominates the other interactions
resulting in a cluster phase (Phase X). The cluster phase is denoted
as Phase X in the phase diagram in Fig.~\ref{fig:a25b05phases}. Note
that the power long-range repulsion caused by stray fields still plays
an important role in causing clusters to order into a hexagonal
``super-lattice", which is illustrated in Fig.~\ref{fig:a25b05snapshots}(j) for
$\rho = 0.25$, $\xi/\lambda=5.0$ and $N_v=2958$. The value of $\xi_c$
is dependent on the density, increasing from $\xi_c=3.3$ at very small
densities to $\xi_c=4.3$ at $\rho\ge0.62$.

%%%%%%%%%%%%%%%%%%%%%%%%%%%%%%%%%%%%%%%%%%%%%%%%%%%%%%%%%%%%%%%%%%%%
% recovered since I think it's importanct to explain why it has attractive well, but has not form cluster in the higher density.  
%%%%%%%%%%%%%%%%%%%%%%%%%%%%%%%%%%%%%%%%%%%%%%%%%%%%%%%%%%%%%%%%%%%%
%Above $\xi_c$, the attractive interaction will be very large in the
%potential(see Fig.~\ref{fig:a25b05allpotentials}). So all of the
%vortex will collaborate together to form the clusters.
For $4.3 \leq \xi / \lambda \leq 10.0$, the ground state of the vortex
system is the cluster. For $3.4 < \xi < 4.3$, the potentials have an
attractive well, but the final configuration of vortex is not the cluster for
large densities ($\rho \ge 0.62$). The system structure is strongly
affected by multiple repulsive length scales.
%The reason is due to that the attractive interaction is too weak to
%force the vortex to form the cluster.  For example, at the $\xi /
%\lambda = 4.0$, for density larger than $0.12$, the system doesn't
%form cluster. But
For low densities $0.05 < \rho \leq 0.12$, the weak attractive
interaction is important and the vortices can form a cluster.
% under this very weak attractive interaction.  Since in the very low
%density, the distance between vortices is very large, so the vortex
%cannot feel the short range strong repulsive interaction. The weak
%attractive interaction can let the vortices form the clusters.
When the density increases, the distance between vortices
decreases, and the short strong repulsive interaction will dominate
and force vortices to form the normal vortex lattice.
% This explains the phase transition line between ordered lattice
% phases and cluster.
%%%%%%%%%%%%%%%%%%%%%%%%%%%%%%%%%%%%%%%%%%%%%%%%%%%%%%%%%%%
%% end of recovery. 
%%%%%%%%%%%%%%%%%%%%%%%%%%%%%%%%%%%%%%%%%%%%%%%%%%%%%%%%%%%

The cluster phase exhibits a nontrivial dependence on the system
size. If number of vortices in the simulation is small, then the
ground state of the system will be a single cluster, shown in
Fig.~\ref{fig:a25clusterCombine} for $\rho=2.37$ and
$\xi/\lambda=9.0$. Note that the RDF for a ground state with multiple
clusters [Fig.~\ref{fig:a25b05rdfs}(j)] and structure factor
[Fig.~\ref{fig:a25b05sf}(j)] is the same as the structure for a single
cluster (Fig.~\ref{fig:a25clusterCombine}). If number of vortices in
the simulation is large enough, multiple clusters will be the ground
state, i.e. there is a maximum size of one cluster. For example, at
$\rho=0.25, \xi/\lambda = 5.0$, the system with 780 vortices will form
one cluster, but the system with 2958 vortices will form
multiple clusters phase.

Here the interior of each cluster is densely packed into a hexagonal
lattice with the density decreasing as one moves outward toward the
edge of the cluster, which is a single ring of vortices equally
spaced. However, the interior structure of the cluster can exhibit
voids, stripes, and other complex
structures.~\cite{varney_hierarchical_2013}

Near the $\xi_c$ boundary, if the system size is too small, the system
will instead feature a large, non-uniform stripe phase. Here we
illustrate a typical vortex configuration, RDF, and structure factor
in Fig.~\ref{fig:a25clusterMiddle} for $\rho=2.37$, $\xi/\lambda=4.75$
and $N_v=2958$. The RDF of middle phase is the same as the RDF of
cluster phase in Fig.~\ref{fig:a25b05rdfs}(j). And $S(\mathbf{k})$
pattern is similar to the multiphase cluster phase in
Fig.~\ref{fig:a25b05sf}(j). The interesting result is that when the
vortices number increases to $N_v=4012$, the whole system will form a
big single cluster. If the vortices number is too small, such as
$N_v=986, 2016$, the ground state of the whole system will be the same
as the low $\xi$ case. For example, at $\rho=2.37, \xi/\lambda=4.75$,
if $N_v=2016$, the ground state of system is kagome lattice(snapshot
is shown in Fig.~\ref{fig:a25b05snapshots}(i)) which is the same as
$\xi<\xi_c$ case. When $N_v=2958$, it forms the cluster middle phase
in Fig.~\ref{fig:a25clusterMiddle}(a). Once the system size reaches
$N_v=4012$, the ground state of the system becomes a single cluster
shown in Fig.~\ref{fig:a25clusterCombine}(a) and increasing the system
size further results in multiple clusters.   This occurs due to long-range repulsive interaction
mediated by stray fields.

There is additional area of the phase diagram where behavior for the
very small density $\rho=0.05$ is not reproducible for larger
densities. At $\rho=0.05$, as $\xi/\lambda$ increases, the system
transforms from a hexagonal lattice to a void phase at
$\xi/\lambda=3.3$ to 3.4, illustrated in
Fig.~\ref{fig:a25clusterVoid}. When the number of vortices is
increased, the system instead forms the cluster middle phase and
cluster phase described above.

\begin{figure*}[!t]
  \includegraphics[width=\textwidth]{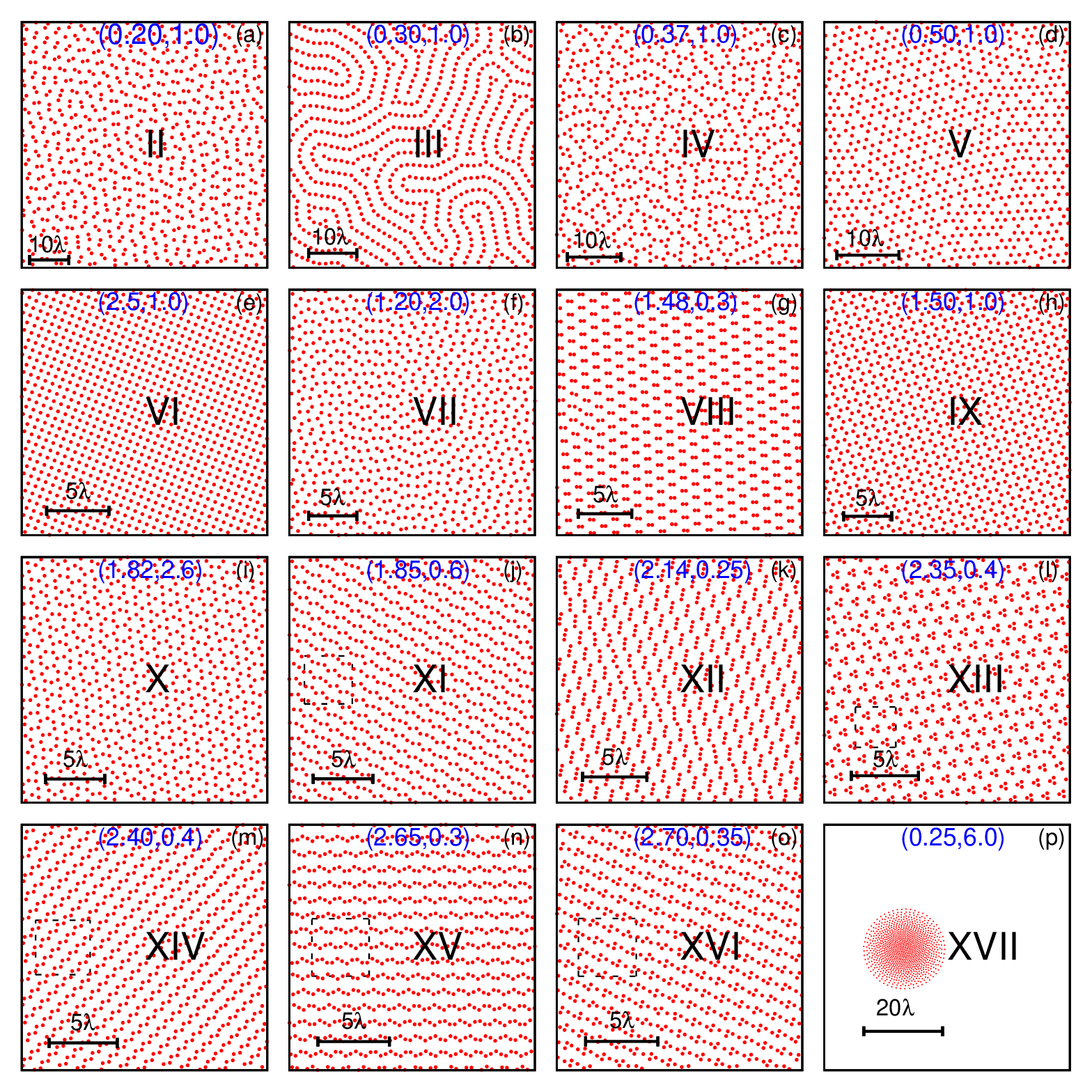}
  \caption{(Color online) Snapshots of a representative vortex
    configuration for each phase (indicated by Roman numerals) in
    Fig.~\ref{fig:a30b15phases}.
    \label{fig:a30b15snapshots}
  }
\end{figure*}

\begin{figure}[!t]
  \includegraphics[width=\columnwidth]{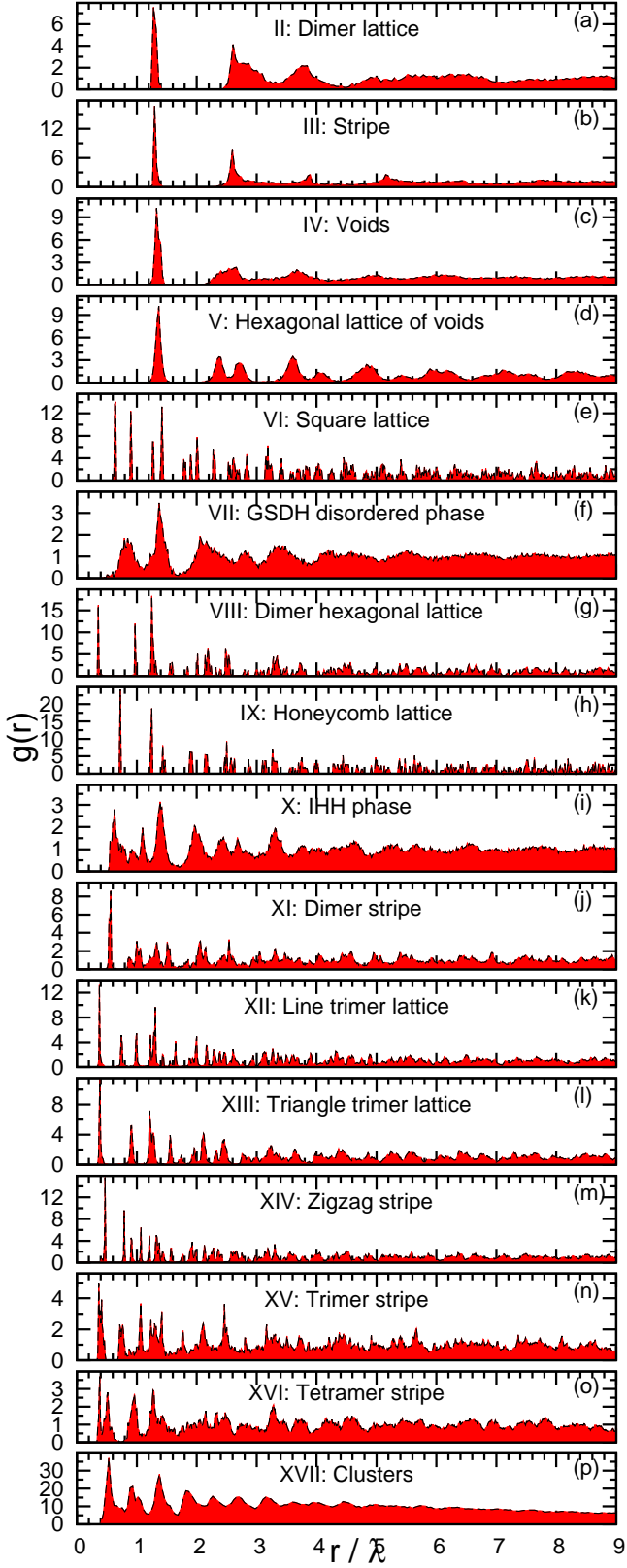}
  \caption{(Color online) Radial distribution function $g(r)$ for
    phases illustrated in Fig.~\ref{fig:a30b15snapshots}.
    \label{fig:a30b15rdfs}
  }

\end{figure}
\begin{figure*}[!t]
  \includegraphics[width=\textwidth]{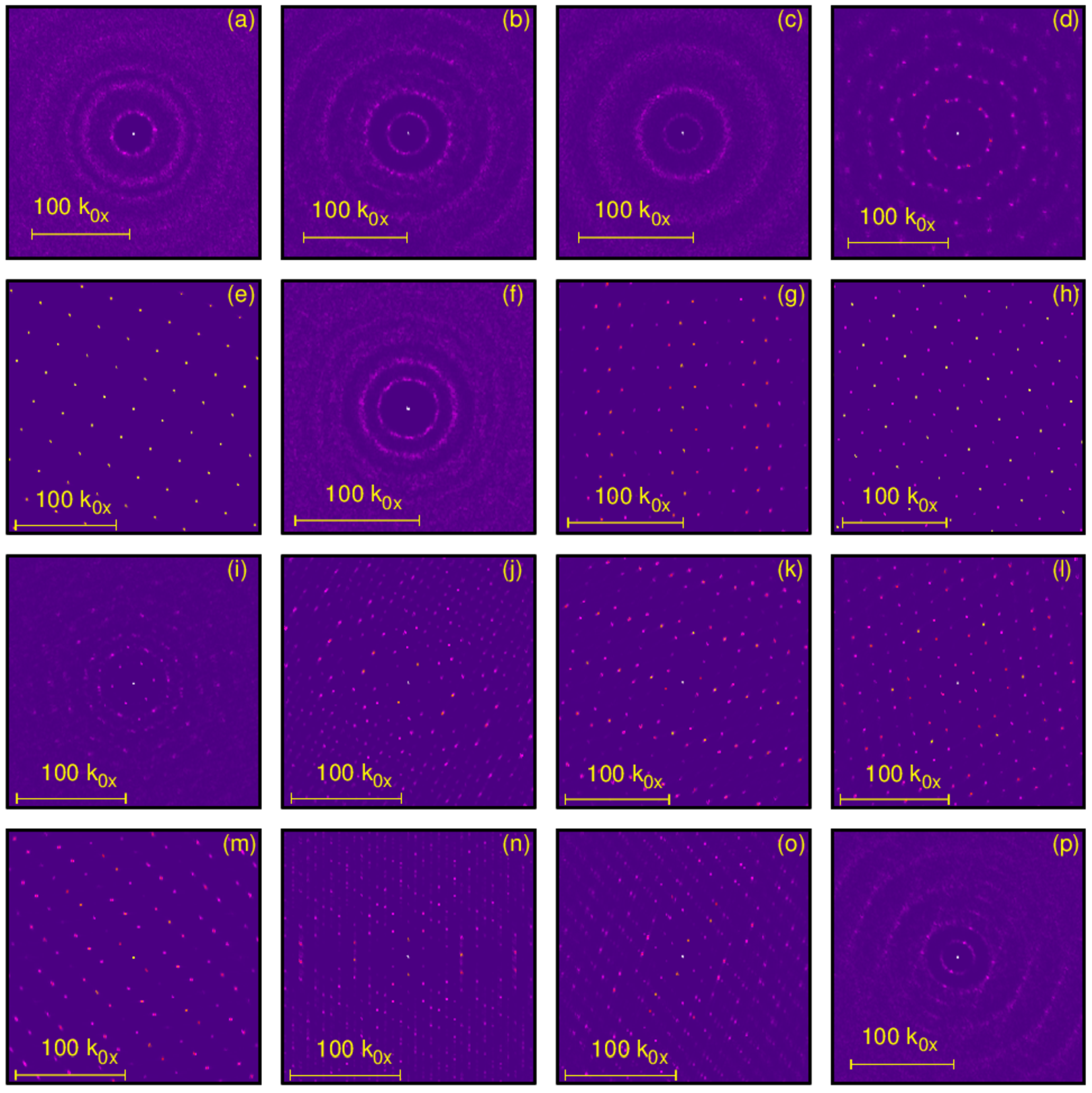}
  \caption{(Color online) Static structure factor $S({\bf k})$ for
    phases illustrated in Fig.~\ref{fig:a30b15snapshots}.
    \label{fig:a30b15sf}
  }
\end{figure*}

\subsection{Phase Diagram II}
% with interaction in Fig.~\ref{fig:a30b15allpotentials}}
The ground state phase diagram in the $\rho$-$\xi/\lambda$ plane
corresponding to Fig.~\ref{fig:a30b15allpotentials} is shown in
Fig.~\ref{fig:a30b15phases}. The main difference with the potential of
Fig.~\ref{fig:a25b05allpotentials} is the presence of a small plateau
located between $r/\lambda = 1.0$ and $2.0$, resulting in a complex
phase diagram with 17 phases overall. As in our discussion on the
previous phase diagram, we represent each phase with a colored
symbol. Phase I is the hexagonal lattice (solid red up-pointing
triangle), Phase II is an unaligned dimer lattice (sky blue empty
circle), and Phase III is a stripe (cadet blue solid circle). Phase IV
is a void state (purple solid rhombus) and Phase V is a hexagonal
lattice with voids (solid blue rhombus). Phases VI and VIII are a
square lattice (solid green square) and an aligned dimer lattice
(magenta solid square), respectively, while Phase VII is a disordered
phase separating the two (solid gray down-pointing triangle). Phase IX
is a honeycomb lattice (solid cyan rhombus) and Phase X is an
intermediate phase between the honeycomb and hexagonal lattice phases
(solid gray pentagon). Phase XI is a dimer stripe (solid dark green
circle), Phase XII is a linear trimer hexagonal lattice (solid violet
down-pointing triangle), and Phase XIII is a trimer lattice (solid
cyan down-pointing triangle). Phases XIV, XV, and XVI are the zig-zag
stripe (solid coral pentagon), trimer stripe (solid turqoise diamond),
and tetramer stripe (solid fuschia diamond) phases,
respectively. Finally, Phase XVII is a cluster phase (solid pink
circle).

\begin{figure}[htb]
  \includegraphics[height=0.69\textwidth]{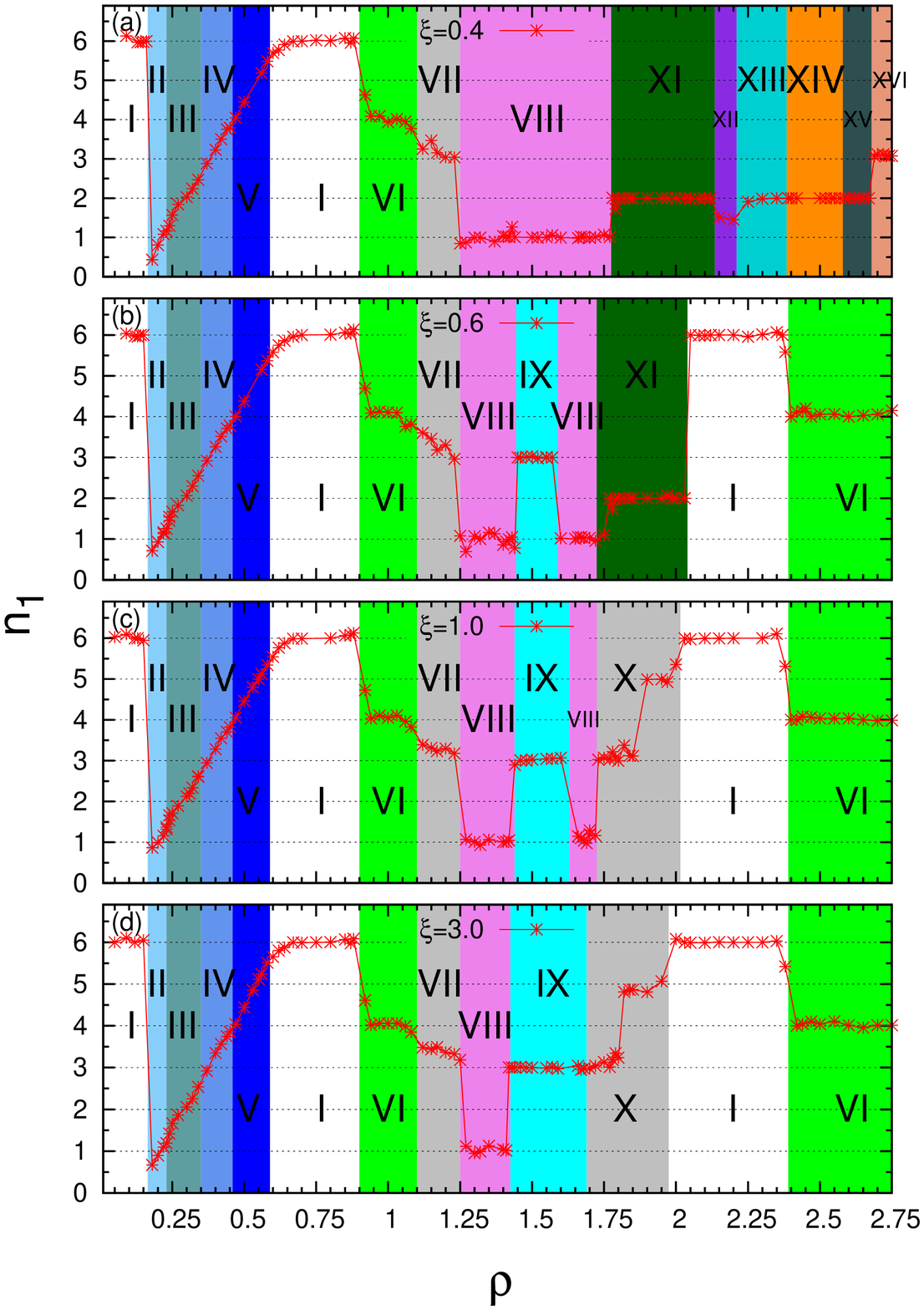}
  \caption{ (Color online) Nearest neighbor number $n_1$ at
    (a). $\xi=0.4$, (b). $\xi=0.6$, (c). $\xi=1.0$, and
    (d). $\xi=3.0$. The background color corresponds to the color used
    for the phase in Fig.~\ref{fig:a30b15phases} and each phase is
    labeled by the appropriate Roman numerals as well.
    \label{fig:n1alla30stl}
  }
\end{figure}

We show the snapshot of a representative final configuration for each
phase in Fig.~\ref{fig:a30b15snapshots}, the corresponding RDF in
Fig.~\ref{fig:a30b15rdfs}, and the corresponding static structure
factor $S(\mathbf{k})$ pattern in Fig.~\ref{fig:a30b15sf}. The number
of the nearest neighbors $n_1$ in the different phases are shown in
Fig.~\ref{fig:n1alla30stl}. With the exception of the hexagonal
lattice phase, note that the background color in
Fig.~\ref{fig:n1alla30stl} corresponds to the color of the point used
in Fig.~\ref{fig:a30b15phases}. Here we show the $n_1$ for $\xi=0.4$,
$0.6$, $1.0$, and $3.0$, which is sufficient to show all of the phase
transitions in Fig.~\ref{fig:a30b15phases}.

First, similar to Phase Diagram I in Fig.~\ref{fig:a25b05phases}, this
system features a critical $\xi_{c}$. When $\xi > \xi_c$, the system
forms giant clusters since the attractive interaction becomes
dominant, with the long-range power law repulsion resulting in
individual clusters forming a hexagonal lattice. Here, $\xi_c$ will
increase from $\xi_c = 3.4$ at $\rho=0.05$ to $\xi_c = 3.75$ at
$\rho=0.46$, then stabilizes to $\xi_c=4.2$ for $\rho\ge0.7$. The
cluster phase is denoted as Phase XVII in the phase diagram in
Fig.~\ref{fig:a30b15phases}. The snapshot, RDF, and static structure
factor of a typical cluster phase are shown in
Fig.~\ref{fig:a30b15snapshots}(p), \ref{fig:a30b15rdfs}(p) and
\ref{fig:a30b15sf}(p), respectively. As with the previous phase
diagram, if the number of vortices is small at fixed $\rho$ and
$\xi/\lambda$, the system will form a single giant cluster which will
become a lattice of clusters as $N_v$ is increased. Similar to Phase
Diagram I, there is also the cluster middle and void phase near the
boundary $\xi_{c}$.

% There is also two types of cluster phase: multiple clusters or only
% one cluster in the system. The violet solid circle represents the
% multiple cluster result and pink solid circle represents the one
% cluster phase in Fig.~\ref{fig:a30b15phases}. All of the results
% were obtained in the $N_v=2958$ system. The snapshot, RDF, and
% static structure factor of a typical cluster phase are shown in
% Fig.~\ref{fig:a30b15snapshots}(p), \ref{fig:a30b15rdfs}(p) and
% \ref{fig:a30b15sf}(p), respectively. Near the boundary from
% $\rho=0.65$ to 2.75, we use much larger vortices system($N_v=4012,
% 5016, 6048, 7020$) to get cluster phase. Similarly as the Phase
% diagram I in Fig.~\ref{fig:a25b05phases}, if the vortices number
% $N_v$ is smaller near the boundary(i.e. $\xi=4.5$), the system will
% form the cluster middle phases(like the phase in
% Fig.~\ref{fig:a25clusterMiddle}(a)) and cannot form a big cluster.

Below $\xi_c$, there are sixteen vortex phases in the phase
diagram. At very low density, $0.05\leq\rho<0.16$, from $\xi = 0.1$ to
$3.4$, the ground state will be a hexagonal lattice (Phase I). For
this phase, the dominant interaction is the long-range repulsive power
law term. Here, the snapshot, RDF, and static structure factor of a
typical configuration are no different from the results shown
previously in Fig.~\ref{fig:a25b05snapshots}(a),
\ref{fig:a25b05rdfs}(a) and \ref{fig:a25b05sf}(a), respectively. Also,
we can confirm that the number of nearest neighbors in this phase is 6
(see Fig.~\ref{fig:n1alla30stl}).

As the density increases, the competition between different length
scales will form an dimer lattice (Phase II) at $\rho = 0.17$ to
$0.23$. The snapshot of a typical configuration is shown for $\rho=0.2$
and $\xi/\lambda=1.0$ in Fig.~\ref{fig:a30b15snapshots}(a). Here we
note that although the vortices pair and form a lattice, the
orientation of each dimer varies throughout the system. Consequently,
the number of nearest neighbors decreases to 1 in this phase and there
is a very sharp beak in $g(r)$ [see Fig.~\ref{fig:a30b15rdfs}(a)]. Due
to the random orientation of the dimers, $S({\bf k})$ has a uniform
ring [see Fig.~\ref{fig:a30b15sf}(a)], reminiscent of the disordered
phases of Fig.~\ref{fig:a25b05phases}.

As the density continues to increase, the dimers will line up and
forming stripes (Phase III). In Fig.~\ref{fig:a30b15snapshots}(b), we
show a typical stripe configuration for at$\rho=0.3$ and $\xi/\lambda
= 1.0$. Because the stripes do not order uniformly, the RDF and static
structure factor do not show strong signatures of Phase IV of
Fig.~\ref{fig:a30b15phases}. In the RDF, the peaks in
Fig.~\ref{fig:a30b15rdfs}(b) are broadened in comparison to
Fig.~\ref{fig:a25b05rdfs}(d). Moreover, the static structure factor
in Fig.~\ref{fig:a30b15sf}(b) has a uniform ring structure instead of
the strong signal in Fig.~\ref{fig:a25b05sf}(d).

Next, the system transitions to a void phase (Phase IV) as a result of
a competition between the repulsive and attractive interactions. A
typical configuration for this phase is shown in
Fig.~\ref{fig:a30b15snapshots}(c) for $\rho=0.37$ and $\xi/\lambda =
1.0$. Here, the system is attempting to form dimers, trimers, and
tetramers but the system is too densely packed to for it to be
possible with $n_1$ varying between 3 and 4 (see
Fig.~\ref{fig:n1alla30stl}). Consequently, the only features visible
in the RDF in Fig.~\ref{fig:a30b15rdfs}(c) are due to short range
pairing and the static structure factor in
Fig.~\ref{fig:a30b15rdfs}(c) has a uniform ring structure.

As density $\rho$ increases from 0.46 to 0.59, the system will form a
hexagonal lattice with voids (HLV) which is denoted as Phase V in
Fig.~\ref{fig:a30b15phases}. Correspondingly, $n_1$ will increase from
4 to 6 which is shown in Fig.~\ref{fig:n1alla30stl}, with $n_1\sim 5$
around density $\rho = 0.55$. The snapshot of the HLV phase is shown
in Fig.~\ref{fig:a30b15snapshots}(d). 
%From the snapshot, we can see
%that the vortices forms a hexagonal lattice, but number of vortices is
%insufficient to produce a full hexagonal lattice. 
As the density
increases, the number of the empty sites will decrease until the
system transforms to a hexagonal lattice. The corresponding RDF and
static structure factor $S(\mathbf{k})$ of HLV is shown in
Fig.~\ref{fig:a30b15rdfs}(d) and \ref{fig:a30b15sf}(d). Note that
$g(r)$ for this phase is consistent with the hexagonal lattice shown
in Fig.~\ref{fig:a25b05rdfs}(a) and there is a six-fold symmetry in
$S(\mathbf{k})$.

From Fig.~\ref{fig:n1alla30stl}, we can find that there is not a very
sharp difference in $n_1$ between dimer, stripe, voids, and
HLV phases. Their nearest neighbor number $n_1$
increases continuously from 1 to 6 and the order of the phases in this
part of the phase diagram is consistent with what is expected for a
system exhibiting a purely repulsive
interaction.\cite{malescio_stripe_2003}
 
Increasing the density further results in a transition from the
hexagonal lattice to a square lattice (Phase VI), where $n_1=4$. A
typical configuration is shown in
Fig.~\ref{fig:a30b15snapshots}(e). The corresponding RDF and static
structure factor for the square lattice are shown in
Fig.~\ref{fig:a30b15rdfs}(e) and \ref{fig:a30b15sf}(e),
respectively. In both $g(r)$ and $S({\bf k})$, the peaks are very
sharp, indicating that there is very little disorder in the system.

Next, the system transitions to a disordered phase (Phase VII) from
$\rho=1.10$ to $\rho=1.25$ due to the competing interactions. Since
this phase separates the square lattice and a dimer hexagonal lattice,
we shall refer to this phase as GSDH. A typical snapshot and
corresponding RDF and $S({\bf k})$ are shown in
Fig.~\ref{fig:a30b15snapshots}(b), \ref{fig:a30b15rdfs}(f) and
\ref{fig:a30b15sf}(f), respectively. As with the previous disordered
phases, the features in $g(r)$ are very broad, indicating a lack of
local structure, and the structure factor has a uniform ring with no
broken symmetries.

From $\rho=1.25$ to 1.43, the ground state of system will
form a dimer hexagonal lattice (Phase VIII). The snapshot, RDF and
static structure factor $S(\mathbf{k})$ for this phase at $\rho =
1.48$ and $\xi/\lambda = 0.3$ are shown in
Fig.~\ref{fig:a30b15snapshots}(g), \ref{fig:a30b15rdfs}(g), and
Fig.~\ref{fig:a30b15sf}(g). Unlike Phase II, the dimers lattice in
Phase VIII have a universal polarization, resulting in sharper peaks
in $g(r)$. Contrasting with Fig.~\ref{fig:a25b05sf}(c), which has a
six-fold symmetry in $S({\bf k})$, the static structure factor of
Phase VIII features a broken symmetry along the $k_x$ direction.

As the density $\rho$ continues to increase, the ground state of
system depends on $\xi/\lambda$ and can be split into two regions. The
first region has both large and small $\xi$ while the second has
intermediate $\xi$ from $\xi\approx0.2$ to $\xi\approx0.5$. For
convenience, we will denote these as Regions I and II,
respectively. 

Let us first consider the transitions in Region I. As the density is
increased while in the dimer hexagonal lattice phase (Phase VIII), the
distance between dimers gradually decreases until the dimer-dimer
distance is equal to the distance between vortices inside a dimer,
resulting in a honeycomb lattice (Phase IX). The snapshot, RDF and
static structure factor $S(\mathbf{k})$ of this phase for $\rho=1.50$,
$\xi/\lambda=1.0$ is shown in Fig.~\ref{fig:a30b15snapshots}(h),
\ref{fig:a30b15rdfs}(h) and \ref{fig:a30b15sf}(h), respectively. The
nearest neighbor number $n_1$ for Region I is shown in panels (b),
(c), and (d) of Fig.~\ref{fig:n1alla30stl} and is consistent with the
expected value of $n_1=3$.

 Increasing the density further results in an intermediate phase
%separating the honeycomb lattice with a reemergent hexagonal lattice
%phase. This intermediate phase
 (IHH for short) is denoted as Phase X
in Fig.~\ref{fig:a30b15phases}. A typical configuration is shown in
Fig.~\ref{fig:a30b15snapshots}(i) for $\rho=1.82$ and
$\xi/\lambda=2.6$. It is immediately clear that there is no local
structure, which is confirmed by the lack of sharp peaks in the RDF
[see Fig.~\ref{fig:a30b15rdfs}(i)]. However, the phase does retain a
six-fold symmetry in $S({\bf k})$ instead of the ring structure one
would expect for a disordered/glassy phase. As the density in this phase is
increased, the number of nearest neighbors increases from $n_1=3$ to
$n_1=5$.

As mentioned previously, Phase X transitions to a hexagonal lattice
(Phase I) as the density is increased. 
%At this point, the vortices are
%so tightly packed that the dominant form of the interaction is the
%short-range repulsive term. 
Increasing the density further results
once again in a transition to a square lattice (Phase VI).

\begin{figure}[tb]
  \includegraphics[height=0.47\textwidth]{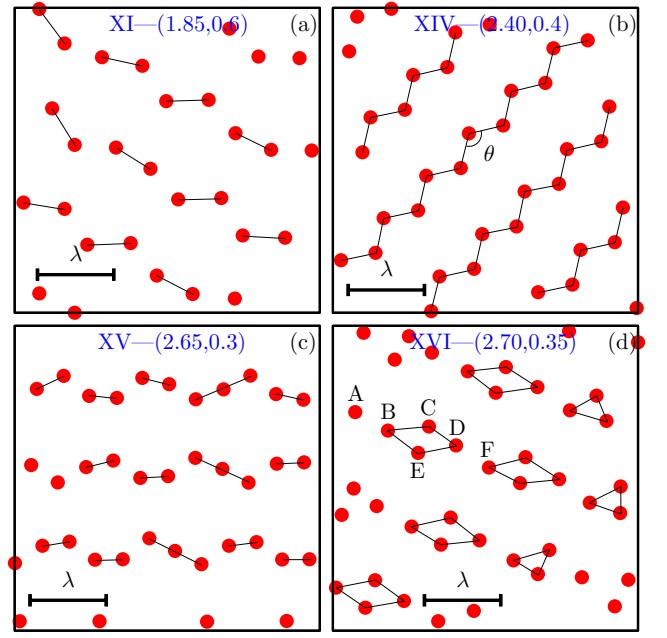}
  \caption{ (Color online) The enlarged dash box in: (a) Phase XI in
    Fig.~\ref{fig:a30b15snapshots}(j); (b) Phase XIV in
    Fig.~\ref{fig:a30b15snapshots}(m); (c) Phase XV in
    Fig.~\ref{fig:a30b15snapshots}(n); (d) Phase XVI in
    Fig.~\ref{fig:a30b15snapshots}(o). We use the black line to link the
    vortices to show the micro-structure of the phase.
    \label{fig:microa30}
  }
\end{figure}

In Region II, the dimer hexagonal lattice (Phase VIII) transitions to
a dimer stripe (Phase XI) phase, which is depicted in
Fig.~\ref{fig:a30b15snapshots}(j) for $\rho=1.85$ and
$\xi/\lambda=0.6$. In this phase, $n_1=2$ and the orientation of the
dimers is not consistent. This can be clearly seen in
Fig.~\ref{fig:microa30}(a), which is an enlargement of the dashed box
in the configuration of Fig.~\ref{fig:a30b15snapshots}(j). The RDF for
this configuration is shown in Fig.~\ref{fig:a30b15rdfs}(j) and
features a sharp peak in connection with the pairing, while subsequent
peaks are broadened due to the varying orientation of the dimers. The
static structure factor [Fig.~\ref{fig:a30b15sf}(j)] exhibits a
two-fold symmetry.

Next, the system transitions to a linear trimer lattice (Phase
XII). A typical configuration is shown in
Fig.~\ref{fig:a30b15snapshots}(k) for $\rho=2.14$ and
$\xi/\lambda=0.25$. Unlike the dimer lattice phase, each linear trimer
is aligned with its neighbors. Consequently, the features in the RDF
and static structure factor are much sharper, as can be seen in
Figs.~\ref{fig:a30b15rdfs}(k) and \ref{fig:a30b15sf}(k),
respectively. Additionally, the vortices in each trimer do not have
the same number of nearest neighbors, with the vortex in the center
having 2 neighbors while the vortices on the ends only having one,
resulting in $n_1\approx4/3$ [see Fig.~\ref{fig:n1alla30stl}(a)].

\begin{figure}[tb]
  \includegraphics[height=0.23\textwidth]{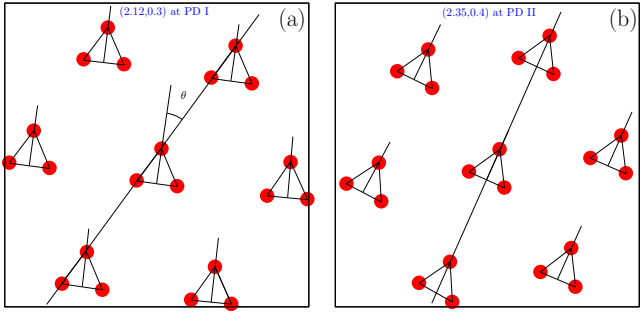}
  \caption{ (Color online) Comparison of the trimer lattices in found
    in (a) Fig.~\ref{fig:a25b05phases} (Phase VII) with $\rho=2.12$
    and $\xi/\lambda=0.3$ and (b) Fig.~\ref{fig:a30b15phases} (Phase
    XIII) with $\rho=2.35$ and $\xi/\lambda=0.4$. The size of this box
    is $3\lambda \times 3 \lambda$. We illustrate individual trimers
    with a solid black line and indicate the orientation of each
    trimer. In panel (a), $\theta=\frac{\pi}{6}$, the trimers are
    aligned to the midpoint between neighboring trimers while in panel
    (b), $\theta=0$, the dimers are aligned to neighboring trimers.
    \label{fig:trimerCompare}
  }
\end{figure}

The next transition is to aligned triangular trimer lattice (Phase
XIII). A representative configuration is shown in
Fig.~\ref{fig:a30b15snapshots}(l). Here, we contrast this phase with
the triangle trimer lattice of Phase Diagram I (Phase VII in
Fig.~\ref{fig:a25b05phases}), showing a close up view of
Fig.~\ref{fig:a25b05snapshots}(g) in Fig.~\ref{fig:trimerCompare}(a)
and a close up view of Fig.~\ref{fig:a30b15snapshots}(l) in
Fig.~\ref{fig:trimerCompare}(b). The size of dashed box in
Fig.~\ref{fig:a25b05snapshots}(g) and
Fig.~\ref{fig:a30b15snapshots}(l) is $3\lambda \times 3 \lambda$. The
main difference between the two phases is the alignment of the
trimers, which is illustrated with a solid black line (there are three
equivalent ways to define the orientation axis for a trimer). The
trimer of Phase Diagram I is aligned to the midpoint between
neighboring trimers while the trimer of Phase XIII in Phase Diagram II
is aligned to neighboring trimers.  Since there is a $\theta=\pi/6$
angle between the orientation of trimer with the alignment direction
of trimers in Fig.~\ref{fig:trimerCompare}(a), we called this phase
polarized triangle trimer lattice. Note that the RDF and static
structure factor of both phases are similar and the number of nearest
neighbors is $n_1=2$. Note that  the aligned triangle trimer lattice
is similar to states of
 colloidal molecular crystal in the presence of 
periodic substrate.~\cite{reichhardt_novel_2002, brunner_phase_2002, agra_theory_2004, sarlah_melting_2005, PhysRevE.85.051401} Another
phenomenon is that the linear trimer and aligned triangle
trimer lattice are extremely close in energy and are nearly
degenerate. In the region with $2.40 < \rho < 2.45$ and $0.25 <
\xi/\lambda < 0.40$, the linear trimer has an energy that is less than
$0.01\%$ smaller than the triangular trimer lattice.

% can both be the ground state of the vortex system at density
% $\rho=2.40$ to 2.45 and $\xi/\lambda=0.25$ to 0.4. This is shown an
% the triangular section between Phase XIII and XIV in the phase
% diagram II in Fig.~\ref{fig:a30b15phases}. In that section,
% different random initial configuration can get different ground
% state: some of them be line trimer lattice and some of them be
% aligned triangle trimer lattice. And the energy of the line trimer
% lattice is $0.00378\%$ smaller than the aligned triangle trimer
% lattice. That's why we mark them as line trimer phase using violet
% down-pointing triangular in the phase diagram II in
% Fig.~\ref{fig:a30b15phases}.

This phenomenon, different initial configuration can get different
stable state at the same condition, happens at several other
places. For example, at $\rho=2.75$, $\xi/\lambda=0.25$ and 0.30, the
tetramer stripe(Phase XVI in Fig.~\ref{fig:a30b15phases}) and trimer
stripe(Phase XV in Fig.~\ref{fig:a30b15phases}) can both be the stable
state based on different initial configuration. At $\xi/\lambda=0.25$,
the energy of trimer stripe is $0.00103\%$ smaller than the tetramer
stripe.

Next, the system transitions to a zig-zag stripe phase (Phase XIV). A
typical configuration is shown in Fig.~\ref{fig:a30b15snapshots}(m)
for $\rho=2.40$ and $\xi/\lambda=0.4$. In this zig-zag phase, there
are two vortices between each bend and the number of nearest neighbors
is $n_1=2$. A close up view of this phase is shown in
Fig.~\ref{fig:microa30}(b) and the angle between nearest neighbours was
calculated to be $\theta\approx116^{\circ}$. The RDF and static
structure factor for this phase are shown in
Fig.~\ref{fig:a30b15rdfs}(m) and \ref{fig:a30b15sf}(m).

Increasing the density further results in a transformation to a trimer
stripe phase (Phase XV). The snapshot, RDF, and static structure
factor $S(\mathbf{k})$ for this phase at $\rho=2.65$ and $\xi/\lambda
= 0.3$ are shown in Fig.~\ref{fig:a30b15snapshots}(n),
\ref{fig:a30b15rdfs}(n) and \ref{fig:a30b15sf}(n), respectively. The
nearest neighbor number $n_1=2$ in Fig.~\ref{fig:n1alla30stl} and we
illustrate a section of the phase in Fig.~\ref{fig:microa30}(c). Here,
each stripe is comprised of both dimers and linear trimers to form
zig-zags of varying depth. The structure factor of this phase also
features a two-fold symmetry like the dimer stripe case in
Fig.~\ref{fig:a30b15sf}(j).

The final phase in Region II is a tetramer stripe (Phase XVI). The
snapshot, RDF, and static structure factor for the the tetramer stripe
at $\rho = 2.70$ and $\xi/\lambda = 0.35$ are shown in
Fig.~\ref{fig:a30b15snapshots}(o), \ref{fig:a30b15rdfs}(o), and
\ref{fig:a30b15sf}(o), respectively. Here, each stripe is mostly
comprised of tetramers with the occasional triangular trimer. The RDF
for this phase features two peaks that are very close in distance and
$S(\mathbf{k})$ features a two-fold symmetry since it's a stripe phase
in the large scale.

In order to clarify the distance between the nearest neighbor in this
phase, we marked six vortices as A, B, C, D, E and F. Vortices B,C,D,E
form a tetramer in Fig.~\ref{fig:microa30}(d). For example, vortex C
as the center, the distance CE is equal to CD and a little bit smaller
than CB. This slightly difference between CE(CD) and CB cause the
first peak of its RDF in Fig.~\ref{fig:a30b15rdfs}(o) forks two very
close peaks. We recognize the first two very close peaks as the first
peak and count the vortices within it as its nearest neighbors. The
nearest neighbor can be recognized as the following: Vortex B's
nearest neighbors are A, C, and E; Vortex C's nearest neighbors are B,
D, and E; Vortex D's nearest neighbors are C, E, and F. Therefore,
every vortex has three nearest neighbors, which we illustrate in
Fig.~\ref{fig:n1alla30stl}(a). 

% The orientation of a trimer and the alignment direction of trimers
% in the polarized triangle trimer lattice is shown in
% Fig.~\ref{fig:trimerCompare}(a) which is the enlarged part of dashed
% box in the panel VII in Fig.~\ref{fig:a25b05phases}. The dashed box
% size is $3\lambda \times 3 \lambda$. Since there is a $\theta=\pi/6$
% angle between the orientation of trimer with the alignment direction
% of trimers in Fig.~\ref{fig:trimerCompare}(a), we call this phase
% polarized triangle trimer lattice.

\section{\label{sec:dis}Discussion}
Layered superconducting systems can have
inter-vortex   
forces with several repulsive and attractive length scales.
We investigated structure formation in a model system with multiple 
repulsive and attarctive length scales.
%There are a variety of new phases in the vortex system with such
%multi-scale length interaction. 
Some of them such as hexagonal, square, honeycomb and
kagome lattice have been reported.\cite{meng_honeycomb_2014} Here we
found other new symmetric phases: dimer hexagonal lattice in
Fig.~\ref{fig:a30b15snapshots}(g), linear trimer lattice in
Fig.~\ref{fig:a30b15snapshots}(k), polarized triangular trimer lattice
in Fig.~\ref{fig:a25b05snapshots}(g) and aligned triangular trimer
lattice in Fig.~\ref{fig:a30b15snapshots}(l). 
% Based on the proposal in
% Ref.~\onlinecite{romero-isart_superconducting_2013}, all of the
% above symmetric phases can be used to trap ultracold atoms.

In addition to conventional lattice phases, we observed a hexagonal
lattice with voids [Fig.~\ref{fig:a30b15snapshots}(d)] and 
various stripe
phases with different unit cells: a dimer stripe
[Fig.~\ref{fig:a30b15snapshots}(j)], a zigzag stripe
[Fig.~\ref{fig:a30b15snapshots}(m)], a trimer stripe
[Fig.~\ref{fig:a30b15snapshots}(n)], and a tetramer stripe
[Fig.~\ref{fig:a30b15snapshots}(o)]. Next, several very stable disordered states 
with no clear local structure were
found in molectular dynamics sumulations. % separating ordered phases.
% Some of glass phases, e.g. GHeS in
%Fig.~\ref{fig:a25b05snapshots}(b), GHoK in
%Fig.~\ref{fig:a25b05snapshots}(f) and GSDH in
%Fig.~\ref{fig:a30b15snapshots}(f), are not only disordered in the
%intermediate and long range scale, but also have no clear local
%structures.
 On the other hand, some of  the very stable phases, e.g. dimer
 phase in Fig.~\ref{fig:a30b15snapshots}(a), stripe in
Fig.~\ref{fig:a30b15snapshots}(b) and voids in
Fig.~\ref{fig:a30b15snapshots}(c), are disordered in the intermediate
and long range scale while having a local structure.

With the same local structure, different arrangements of local
structure can form different phases which will affect the property in
the intermediate and long-range scales. For example, there are several
types of dimer lattices based on different dimer orientation. Dimers
can locally form hexagonal lattice in a domain in the Phase III in
Fig.~\ref{fig:a25b05snapshots}(c). Dimers can also be 
disordered in the Phase II in Fig.~\ref{fig:a30b15snapshots}(a). All
of dimers can have a universal orientation and form a hexagonal
lattice in Phase VIII of Fig.~\ref{fig:a30b15snapshots}(g). Dimers can
also line up to form a stripe
[Fig.~\ref{fig:a25b05dimerrho150}(a)]. 
%The difference in the
%intermediate and long-range scales can be seen from the structure
%factors $S(\mathbf{k})$ shown in Fig.~\ref{fig:a25b05sf}(c),
%\ref{fig:a30b15sf}(a) and \ref{fig:a30b15sf}(g).

Finally, there are three types of trimer lattice with different
arrangement of trimers in this system. Three vortices can form two
types of trimer: linear trimer [shown in
  Fig.~\ref{fig:a30b15snapshots}(k)] and two types of triangluar
trimers [shown in Fig.~\ref{fig:a25b05snapshots}(g)
  and~\ref{fig:a30b15snapshots}(l)]. 
  The angle $\theta$ between the
orientation of triangular trimers and the alignment direction of
triangle trimers can be different in various arrangements. 
In the
Phase VII in Fig.~\ref{fig:a25b05snapshots}(g) [close up view in
 Fig.~\ref{fig:trimerCompare}(a)], $\theta=\pi/6$, it formed a
polarized triangular trimer. In the Phase XIII in
Fig.~\ref{fig:a30b15snapshots}(l) [close up view in
 Fig.~\ref{fig:trimerCompare}(b)], $\theta = 0$, it formed an aligned
triangular trimer lattice.

\section{\label{sec:summary}Summary}
In conclusion, quantum vortices in superconductors may offer a unique
route for engineering magnetic field configurations required for
quantum emulators. This application demands creation of various
geometries of vortex matter which is not possible in type-2
superconductors in the absence of pinning center.

Here we propose to utilize type-1.5 superconductors which possess
several attractive and repulsive length scales in the intervortex
interaction potential. Several repulsive lengths scales can be
engineered in layered structures, where difference layers have
different magnetic penetration lengths or in superconductor-insulator
multilayers where magnetic field lines can spread in the insulating
layers due to suppression of the Meissner effect.

We studied this situation utilising a  model of point-particles with
effective interaction potential. This model neglects non-pairwise
intervortex forces, which in fact are small under certain
conditions.~\cite{PhysRevB.84.134515} In case of the realization in
layered system the applicability of the point-particle model assumes
high vortex line tension which makes the system translationally
invariant in $z$-direction. We demonstrated that in such systems it is
possible to realize a wide range of  vortex states. Recent
study of the vortex cluster phase
in type-1.5 superconductors, similar to
the ones considered here, demonstrated that
the disordered vortex clusters states have glass dynamics.\cite{diazmendez_glassy_type15}
%CITE  R. Diaz-Mendez et. al. to be published.} 

\section*{\label{sec:ack}Acknowledgements}
This work was supported by the National Science Foundation under the
CAREER Award DMR-0955902, Goran Gustafsson Foundation  and by the Swedish
Research Council 642-2013-7837. Q.~M. wants to thank the inspired discussion with
Gregory M. Grason.

\bibliography{phaselong7}

\begin{thebibliography}{45}
\bibitem [{not()}]{note1}%
  \BibitemOpen
  \href@noop {} {}\bibinfo {note} {The hexagonal lattice order $a$ can be
  calculated given the density $\rho$. The relation between them is
  $\rho=(\sqrt{3}a^2/2)^{-1}$. Then we can get $a=\sqrt{2/(\sqrt{3}\rho)}$. If
  $\rho=1.12$, then $a=1.02$; $\rho=1.18$, then $a=0.99$.}\BibitemShut {Stop}%
\end{thebibliography}
\end{document}